\documentstyle[prb,aps]{revtex}

\begin{document}
\draft
\title{Bilayers of Chiral Spin States}
\author{Carlos R.~Cassanello and Eduardo H.~Fradkin}
\bigskip
\address
{Loomis Laboratory of Physics and Materials Research Laboratory\\
University of Illinois at Urbana-Champaign\\
1110 W.Green St., Urbana, IL, 61801-3080\\
}

\maketitle

\begin{abstract}
We study the behavior of two planes of Quantum Heisenberg
Antiferromagnet in the regime in which a Chiral Spin Liquid is stabilized
in each plane. The planes are coupled by an exchange interaction of
strength $J_3$.
We show that in the regime of small $J_3$ (for both
ferromagnetic {\it and} antiferromagnetic
coupling), the system dynamically selects an
\underline{antiferromagnetic}
ordering of the ground state {\it chiralities} of the planes.
For the case of an antiferromagnetic interaction between the planes, 
we find that, at some critical value $J_3^c$ of the inter-layer coupling, 
there is a phase transition to a valence-bond state on the interlayer links.
We derive an effective Landau-Ginzburg theory for this phase transition. It
contains two $U(1)$ gauge fields
coupled to the order parameter field. 
We study the low energy spectrum of each phase. 
In the condensed phase an ``anti-Higgs-Anderson" mechanism occurs. 
It effectively
restores time-reversal invariance  
by rendering massless one of the gauge fields
while the other field locks the chiral degrees of freedom locally.
There is no phase transition for ferromagnetic couplings.
\end{abstract}




\section{Introduction}
\label{sec:intro}

The discovery of superconductivity at high temperatures in the otherwise
insulating copper oxides has motivated a thorough search for new
physical mechanisms for both superconductivity and antiferromagnetism.
This search has produced a host of new possible mechanisms many of which
are not yet established on solid ground. Among these new ideas,
the anyon mechanism\cite{kla} stands as, perhaps, the most novel of
them. For this
reason, it has attracted a lot of attention. At a microscopic level, the
anyon state requires that the underlying insulating state, known as the
Chiral Spin Liquid\cite{wwz} (CSL), should
necessarily break Time Reversal ($T$) invariance and Parity ($P$). 
An experimental
signature of a state with broken $T$ and $P$ invariance is optical
dichroism~\cite{prediction}.
So far, however, there is no
experimental evidence in support of the spontaneous breaking of either
$T$ or $P$ in the copper oxides~\cite{experiments}. 
Clearly, the simplest option is that
these symmetries are not broken in the copper oxides and that the
insulating states are unrelated to the CSL. At the present time this
appears to be the case.

In this paper we will explore the possibility that $T$ and $P$
may be broken in one individual plane but not on the system as a whole.
Individual isolated planes may still be in states which break $T$ and
$P$ but the {\it sign} of this breaking may not be the same from plane
to plane.
The simplest case is to imagine that the copper oxide planes are coupled
by some interaction and that this coupling is responsible for the
selection of the state. A version of this problem has been studied by
Rojo and Leggett\cite{rojo}. They considered two planes with a {\it
doped} CSL on each plane and, hence, had an {\it
anyon superconductor} on {\it each plane}. They further assumed that
the planes were coupled together only by a direct Coulomb
interaction between the anyons on each plane. They did not fix {\it a
priori} the relative sign of the statistics
of the anyons on each plane but, instead, asked which {\it relative
sign} was preferred by the Coulomb interactions. They found that the
Coulomb interactions prefer the relative statistics to  be {\it
antiferromagnetic} ordered, namely opposite signs. The Rojo-Leggett
result is due to a rather subtle edge effect. In fact, they found no
effect in the bulk.

In many copper oxides, the physical situation is such that the
planes come in groups in which  the planes are closer together than
among nearby groups. This is rather common in the Bismuth based copper
oxides. Because in these materials the {\it inter-layer} exchange
constant which couples the copper spins can be comparable to the {\it
intra-layer} exchange constant, there is a competition between {\it
intra} and {\it inter} layer types of ordering. 
Quite generally, 
one expects to find to distinct regimes in the phase diagram for bilayers.
At weak interlayer coupling, 
the ground state of the individual layers may be stable.
However, if the interlayer exchange coupling dominates, 
the likely ground state should
be a valence bond state  on the interlayer links.  
The case of two coupled
Ne{\'e}l states was considered recently by Uhbens and Lee\cite{lee}, by
Millis and Monien\cite{millis} and by Sandvik and 
Scalapino\cite{scalapino}. These authors considered the effects of an  
inter-layer exchange interaction on the Ne{\`e}l ground states of the planes. 

In this paper we will reconsider the problem of
a bilayer of quantum antiferromagnets in a regime in which there is
enough frustration to drive each plane separately into a Chiral Spin
Liquid. The planes will be assumed to be
coupled by an antiferromagnetic exchange interaction of strength $J_3$.
The problems that we want to address are: (a) does the inter-layer
exchange interaction select the relative ordering of the chiralities and
(b) what is the phase diagram for this system as a function of the
inter-layer interaction. 
We consider a situation in which there is  a CSL ground
state on each plane, with fixed chirality but arbitrary sign.
We find that quantum fluctuations around this state select an
 {\it antiferromagnetic}
ordering of the chiralities. This is a rather interesting result. It
means that even if on each plane the system was allowed to break $P$ and
$T$, the dynamics selects the state which is on the whole $P$ {\it and}
$T$ invariant. We also find that, as $J_3$ increases, there is phase
transition to a state that we identify as a valence bond state  on the
inter-layer links, namely a $T$ and $P$ invariant spin gap state very
similar to the one found by Ubbens {\it et al}\cite{lee}, Millis {\it et
al}\cite{millis} and Sandvik {\it et al}\cite{scalapino}.
The problem of the ordering of chiralities by an
inter-layer exchange interaction was considered previously by Gaitonde
{\it et al}\cite{gaitonde}. By means of a perturbative expansion in
powers of $J_3$ they concluded that the chiralities order {\it
ferromagnetically}. The results that we report here disagree with those
of Gaitonde {\it et al}. 

As it is by now well known\cite{afm,kotliar,wwz,book}, the CSL state and
its low-lying
excitations can be described in terms of an effective continuum field theory
which is very much analogous to a set of Dirac self-interacting fermions
in two space and one time dimensions. We find that the essential physics
of this system can be understood in terms of the properties of an
effective continuum theory of Dirac fermions on each plane provided that
a physically sensible cutoff is introduced. 
The effective model contains two sets of
massive Dirac fermions on each plane. The chirality of the state is given 
by the sign of the mass term. As in WWZ, the
fluctuations around the CSL of each plane 
are represented by gauge fields (one for each plane). 
By a detailed microscopic analysis we find that the interlayer exchange
fluctuations are represented by a {\it complex} order parameter field. 
The effective theory is controlled by three parameters: 1) the magnitude of the
fermion mass on each plane ({\it i.e.,\/} the fermion gap in the CSL), 2) the
interlayer exchange constant 
(which determines the energy gap for fluctuations of
the order parameter) and 3) the number of fermionic species 
(which we take to be $N$). In this picture, the phase
transition to the valence-bond state becomes the phase transition to a state 
in which the
complex order parameter acquires a non-vanishing expectation value.
Our basic strategy is to
first derive this effective theory and then use it to address the issues
of the ordering of chiralities and of the nature of the phase diagram.

Mean-field theories of frustrated antiferromagnets on a single plane have yielded a
host of possible non-magnetic variational ground states. The actual phase diagram is
not known in detail although it is generally accepted that non-chiral states are
somewhat favored by variational calculations. In this paper we will not consider how
interlayer couplings may alter this competition among possible single layer
variational states. Rather, we will describe how interlayer interactions disrupt the
CSL in favor of an interlayer valence bond state, which is clearly favored at strong
coupling. The determination of the global phase diagram for bilayers is an interesting 
problem which is however still outside the reach of present theoretical tools and
beyond the scope of this article.

The effective field theory of fermions can be studied 
within a $1/N$ expansion. We use
this expansion for two different purposes. 
First we look at the quantum corrections 
to the ground state energy of a system in which the two 
CSL are decoupled. We find
that, at leading order in the $1/N$ expansion, 
the state with antiferromagnetic
(opposite) chiralities is degenerate with 
the state with ferromagnetic chiralities.
However, we find that the leading corrections, 
due to fluctuations of interlayer
exchange processes, the state with antiferromagnetic 
ordering of chiralities is selected. 
In addition to the spontaneous breaking of this discrete symmetry 
(the relative chirality), 
the fermionic theory for the bilayers undergoes a dynamical
breaking of the interlayer (out-of-phase) 
gauge symmetry at a critical value of the
interlayer coupling constant. This phenomenon is
strongly reminiscent of the breaking of 
chiral symmetry in the related (but not equal) field theoretic Gross-Neveu 
and Nambu-Jona Lasinio models~\cite{gross}.
Also, within this $1/N$ expansion, 
we find a phase transition from a regime in which
the two planes have CSL ground states with opposite signs, 
to a state in which the
inter-layer order parameter field condenses. 
We further investigate the physics of
this phase transition by deriving an effective 
Landau-Ginzburg-type field theory,
valid in the vicinity of the phase 
transition, {\it i.~e.\/} for $J_3 \sim J_3^c$.

The degrees of freedom of the Landau theory, 
which is fully quantum mechanical,
are the interlayer order parameter field 
and the gauge fields of the two planes. We
present a qualitative
study of the fluctuation spectrum of the two phases. 
The weak coupling phase has
(almost) the same spectrum as that of two 
CSL with opposite chiralities: semions with
opposite chiralities and gapped gauge fluctuations). 
However, a the phase with broken
symmetry (in which the interlayer field condenses) displays an interesting
``anti-Higgs-Anderson" mechanism: 
the condensation of the order parameter field causes a
gauge fluctuation, which is massive in the 
unbroken phase due to the Chern-Simons
terms, to become massless. 
This, in turn, implies that any excitation which couples to
the gauge fields (the semions, in particular) 
to become confined by strong, long
range, logarithmic interactions. 
The resulting spectrum of the condensed phase is {\it
equivalent} to the low-lying spectrum 
of a ground state of local singlets, {\it i.~e.\/}
a valence-bond state on the interlayer links. The interlayer gauge field,
remains massive and it effectively disappears from the spectrum.
Thus, the ``anti-Higgs-Anderson" mechanism  
wipes out all trace of broken time-reversal-invariance in the system.
Unexpectedly, in this phase the system is actually more symmetric than in the 
non-condensed state.

The paper is organized as follows. 
In section \ref{sec:mft} we introduce
the model for the bilayer and develop the mean field theory
and briefly discuss the phase diagram. 
In section \ref{sec:order} we address the
problem of the dynamical selection of chiralities. 
In section \ref{sec:L-G} we derive a gradient expansion for the 
low energy modes of the (two) gauge fields and the relevant (scalar)
channel of the field coupling the planes. 
In section \ref{sec:symm} we discuss the properties of the symmetric
phase where the field coupling the planes does not condense, and
an effective action for the gauge fields is derived and studied.
Section \ref{sec:higgs} deals with the broken symmetry phase.
Section \ref{sec:conc} is devoted to the
conclusions. We also include appendices which contain technical details of
the mapping onto the effective continuum theory and the 
computation of Feynman diagrams relevant for the phase transition,
the ordering of the chiralities and the gradient and $\frac{1}{N}$
expansions.

\section{Mean Field Theory for two coupled Chiral Spin States}
\label{sec:mft}

Our model consists of two square-lattice spin-$\frac{1}{2}$ Heisenberg
antiferromagnets coupled through an exchange interaction of
nearest-neighbors spins between planes with strength $J_3$, and
nearest-neighbors ($J_1$) and next-nearest-neighbors ($J_2$)
interactions on each plane.
The lattice Hamiltonian reads
\begin{equation}
H = H_L + H_U + J_{3} \sum_{\vec x}^{} {\vec S}_L (\vec x) \cdot {\vec S}_U
(\vec x + {\vec e}_z)
\end{equation}
where $H_{L,U}$ is the usual Heisenberg Hamiltonian,
\begin{equation}
H_{L,U} =  J_{1} \sum_{\vec x, j=1,2}^{}
        {\vec S}_{L,U}({\vec x}) \cdot {\vec S}_{L,U}
        ({\vec x} + {\vec e}_j )
+ J_{2} \sum_{\vec x, j=+,-}^{}
        {\vec S}_{L,U}(\vec x) \cdot {\vec S}_{L,U}(\vec x +{\vec e}_1 
+ j{\vec e}_2) 
\end{equation}

Using the slave fermion approach,
the spin operator can be written in terms of fermionic creation
and anihilation operators
${\vec S} (\vec x) \equiv c^{\dag}_{\alpha} (\vec x) {\vec \sigma}^{\alpha\beta}
c_\beta (\vec x) $ with the usual constraint of single occupancy.
We decouple the quartic terms
by using a standard  Hubbard-Stratonovich (H-S) transformation.
Up to an integration over the H-S fields,
the original theory is equivalent to
the one that follows from the action
given by the lagrangian:
\begin{equation}
{\cal L} = {\cal L}_L + {\cal L}_U 
-
{1\over J_3}\sum_{\vec x}^{} |\chi_{z}(\vec x) |^2 
+ \sum_{\vec x}^{}
    \left[ c^{*}_{L}(\vec x) \chi_z (\vec x) 
c_{U} (\vec x +{\vec e}_z) + h.c.\right]
\label{lagrangian}
\end{equation}
where
\begin{eqnarray}
{\cal L}_L & = & \sum_{\vec x}^{}
               c^{*}_{L} (\vec x) \left( i \partial_t + \mu \right)
               c_{L} (\vec x)
             + \sum_{\vec x}^{}
\varphi_L(\vec x) \left( c^{*}_{L}(\vec x) 
c_{L}(\vec x) - 1 \right) \nonumber \\
& - &
{1\over J_1} \sum_{\vec x ; j=1,2}^{} |\chi_{j,L}(\vec x) |^2
-
{1\over J_2}\sum_{\vec x ; j=+,-}^{} |\chi_{j,L} (\vec x) |^2\nonumber \\
& + &\kern-0.35cm
\sum_{\vec x ;j=1,2}^{} \left[ c^{*}_{L}(\vec x) \chi_{j,L} (\vec x) 
c_{L} (\vec x + {\vec e}_j )
+ h.c.\right]
+ \kern-0.35cm \sum_{\vec x ; j=+,-}^{}
    \left[ c^{*}_{L}(\vec x) \chi_{j,L}(\vec x) 
c_{L} (\vec x +{\vec e}_1+j{\vec e}_2)+h.c.\right]
\label{planelag}
\end{eqnarray}
where we have dropped the spin indices  $\alpha, \beta$ to simplify the notation,
with a similar definition for ${\cal L}_U$.
Here $\mu$ is the chemical potential and $\vec x$ means $(\vec r , t)$.
The constraint of single occupancy is enforced
by the bosonic Lagrange multiplier
field $\varphi (\vec x)$.
This type of factorization was originally proposed
by Affleck and Marston \cite{afm} and by Kotliar \cite{kotliar}.
The H-S fields can be parametrized in terms
of an amplitude $\rho_{j}(\vec x)$
and a phase $ A_{j}(\vec x)$.
This Lagrangian has a local symmetry
if the Lagrange multiplier field $\varphi$
transforms as the $ A_0$ component of
a $U(1)$ gauge field.

The MFT consists in integrating out
the fermions, at a fixed density,
and treating the fields $\chi_j (\vec x)$
within a saddle-point expansion.
As it is well known, one serious problem with this mean field theory, is
that there is no small parameter
in powers of which to organize
the semi-classical expansion. Following Affleck and Marston\cite{afm},
we will allow the number of spin species to
run to N instead of 2,
which is the case for the spin-${1\over 2}$ Heisenberg model.
After re-scaling the coupling constant strengths J's
and the fluctuating part of the fields,
a 1-loop expansion of the fermionic determinant
around the $N\to\infty$ Mean-Field solution can be
performed by keeping the diagrams up to order ${1\over N}$.
We have
${\cal S}_{eff}\left[\varphi,\chi_j\right] = 
N \bar{\cal S}\left[[\varphi, \chi_j\right]$,
and the quantum partition function is
${\cal Z} = \int {\cal D} \chi {\cal D} \chi^{*} {\cal D} \varphi 
e^{iN\bar{\cal S}}$.

There exists a whole family of solutions
of the saddle-point equations.
The simplest solutions
are the valence bonds states and the flux phases.
These may or may not be chiral.
In this work we consider the problem of the selection of the relative
chirality of a state in which there is a Chiral Spin Liquid on each
plane. Thus, we {\it choose} a saddle point which represents Chiral Spin
States on each plane and we will investigate which configuration of
chiralities is chosen dynamically.

Wen, Wilczek and Zee\cite{wwz} (WWZ) have given a construction of the
Chiral
Spin State, which was first proposed by Kalmeyer and Laughlin\cite{kla}.
WWZ begin with the flux phases, which have a uniform
value for the amplitude of the n.n. H-S fields, say $\rho(\vec x) = {\bar\rho}$.
This amplitude however, can fluctuate.
The phases of the Bose fields
on the n.n. links of an elementary plaquette
have a circulation equal to $\pi$ or $-\pi$ in mean-field value.
This feature produces a collapse
of the Fermi surface into four discrete
points of the Brillouin zone:
$\left(\pm {\pi \over 2a},\pm {\pi \over 2a}\right)$
at which two bands of states
(positive and negative energy,
``conduction" and ``valence" bands) become degenerate.
At these points,
the excitation spectrum is linear and gapless.
This allows for a mapping onto
a discrete version of the Dirac theory
with two massless fermion species of two-component spinors,
with the ``speed of light"
equal to the Fermi velocity $ v_F = 2a{\bar \rho}$.
This gapless state can become unstable
due to the effects of fluctuations.
Several channels are known to be possible.
If the staggered part of the fluctuations
of the amplitude of the Bose fields on the n.n. links,
picks up a non-zero expectation value,
gaps will open up in the elementary excitation
spectrum and they will provide masses (or gaps)
to the Dirac-like fermionic excitations.
These fluctuations can be seen
to drive the flux phase into a dimer
or Peierls state and do not break time reversal invariance or parity.

A mass term in a Dirac equation
for a {\it single} two-component spinor Fermi field in $2+1$-dimensions
generally breaks T and P since
the Hamiltonian,
while hermitian, becomes complex.
Since all three Pauli matrices are involved
(two for the gradient terms and the third one for the mass term)
there is no basis
in which the Hamiltonian could be real.
Therefore, the Hamiltonian
is not self-conjugate and T is broken.
However, in the case in which two species of fermions are present,
 the presence of such mass terms does not necessarily break P and T
since they may have opposite sign for the different species.
This is the case of the so-called
Peierls mass, which occurs in dimer phases.
It is here where frustration comes to play a crucial role.
By turning on n.n.n. interactions,
WWZ allowed for additional H-S fields on
the diagonals of the elementary plaquettes.
The MF configuration for the phases
can be arranged so that each triangle
in an elementary plaquette
is pierced by a flux equal to $\pi \over 2$.
In this way,
a time-reversal and parity breaking mass
can be generated, {\it i.e.,\/} one can provide a mass
with the {\it same sign} to both fermion species in the plane.
In order
to perform the
mapping onto the Dirac theory
it is necessary to introduce
four different field amplitudes
at each unitary cell of four sites.
This procedure can be done on the real space
lattice by defining four sublattices and
assigning an independent field amplitude
to each one and expanding in gradients of
the field amplitudes\cite{book}, or on the reciprocal
lattice\cite{wwz,thesis} by expanding the lattice amplitude at
each point as a linear combination of
four independent fourier components amplitudes.
On the reciprocal lattice these fields
are the fourier components of the lattice amplitude
centered at the four Fermi points.
The low-energy physics
of the system is determined by the scattering processes
among these four amplitudes.
Any of these procedures is equivalent to a folding
of the first Brillouin Zone.

In the CSS, the mean-field ansatz
for the amplitudes and phases of the H-S
fields on the n.n. and
n.n.n. links 
is given by~\cite{wwz,thesis}
\begin{eqnarray}
\bar{\chi}_1(e,e) & = & - \bar{\chi}_1(o,e) = \bar{\chi}_1(e,o) = 
-{\overline\chi}_1(o,o) = i{\overline\rho}
\nonumber \\
{\overline\chi}_2(e,e) 
& = & - {\overline\chi}_2(o,e) 
= -{\overline\chi}_2(e,o) = {\overline\chi}(o,o) = -i{\overline\rho}
\nonumber \\
{\overline\chi}_{+}(e,e) & = & {\overline\chi}_{+}(o,e) = 
- {\overline\chi}_{+}(e,o) = - {\overline\chi}_{+}(o,o) = 
i{\mathchar'26\mskip-9mu\lambda}
\nonumber \\
{\overline\chi}_{-}(e,e) & = & {\overline\chi}_{-}(o,e) = 
- {\overline\chi}_{-}(e,o) = -{\overline\chi}_{-}(o,o) = 
-i{\mathchar'26\mskip-9mu\lambda}
\end{eqnarray}
The fields $\bar{\chi}_j$, with $j=1,2$ or $j=+,-$
are the H-S fields sitting on the n.n. and n.n.n. links
respectively.
The four different sublattices are denoted by
$(e,e)$, $(o,e)$, $(e,o)$, $(o,o)$, where $e$ and $o$
mean even or odd site respectively.

Once the mean-field H-S ansatz has been used
into the Hamiltonian for one plane,
convenient linear combination of
the four field amplitudes
can be arranged in the form of two {\it two-component}
spinors and one can re-write the lagrangian for
a single plane in the form of a
lattice Dirac lagrangian with two massive
fermion species. So far we did not
include any fluctuations of the H-S fields.
We will be interested in the fluctuating
part of the {\it phase} of the H-S fields.

In order to capture the physics
of the system in the regime of
long-wavelength, low-energy of the spectrum, we
do not need the full lattice theory,
but a linearized version around the Fermi points
that keeps
all the scattering processes that are responsible 
for the behavior of the low-energy
excitations of the system. 
In the case of only one square lattice
bearing a Chiral Spin State, we
arrive to a $2+1$-dimensional effective action
involving two massive relativistic fermions coupled
to a gauge field~\cite{wwz,book,thesis}.
The form of this action is given by
\begin{equation}
{\cal S} = \int dx_0\int dx^2 \left\{ \bar{\psi}_{1}
\left( i 
{\raise.15ex\hbox{$/$}\kern-.57em\hbox{$\partial$}} 
- {\raise.15ex\hbox{$/$}\kern-.72em\hbox{$A$}} - m_1\right) \psi_{1}
+ \bar{\psi}_{2}\left( i 
{\raise.15ex\hbox{$/$}\kern-.57em\hbox{$\partial$}} 
- {\raise.15ex\hbox{$/$}\kern-.72em\hbox{$A$}} - m_2\right)\psi_{2}\right\}
\label{splane}
\end{equation}

The continuum field $\psi_a$ is related
to the lattice amplitude $\Psi_a$ by $\psi_a(\vec x)\equiv \Psi_a(\vec x)/a$.
We use a representation of Dirac gamma
matrices in which
$\gamma_0=\sigma_3$; $\gamma_1=-i\sigma_2$ and
$\gamma_2 =-i\sigma_1$, where $\sigma_j$, $j=1,2,3$
are the usual Pauli matrices.
The coupling to the gauge field (the {\it statistical vector potential} \ )
$A_\mu$ comes through the covariant derivative 
${\raise.15ex\hbox{$/$}\kern-.70em\hbox{$D$}} \equiv
{\raise.15ex\hbox{$/$}\kern-.57em\hbox{$\partial$}} 
- i {\raise.15ex\hbox{$/$}\kern-.72em\hbox{$A$}} $.
The statistical vector potential is given by
$A_j\equiv {\tilde\phi}_j/a=2{\overline\rho}{\tilde\phi}_j/v_F$ and $A_0 
= \varphi/v_F$,
where ${\tilde\phi}_j$ is the fluctuating part
of the phase of the Hubbard-Stratonovich fields
on the n.n. links, $\varphi$ is the Lagrange multiplier 
field~\cite{wwz,book,thesis} and $x_0 \equiv v_F t$.

The masses of the fermions come from the
amplitude of the H-S field on the n.n.n.
links and give a measure of the amount of frustration
present in the system. These masses, although not
necessarily equal in magnitude, have the same sign
for both species.
We assume that these amplitudes are fixed at their
mean field values, since we are interested only in 
the effects of interlayer fluctuations.

In what follows we adapt the methods of 
references \cite{wwz} and \cite{book} to the bilayer problem~\cite{thesis}.
We have a duplication of terms due to the inclusion of the
second plane and new terms arising from the interplanar
interaction.
In the continuum limit, the action for the fermions
in the low energy theory has two species of
Dirac fermions {\it on each plane} coupled to both the {\it intra-layer}
and {\it inter-layer} Hubbard-Stratonovich fields which mediate the
interactions among the fermionic degrees of freedom. For simplicity we
will assume the degree of chiral breaking is fixed and parametrized
by two non-fluctuating masses $m_U$ and $m_L$.
These {\it masses} are given by $m_{L,U}\equiv 
4{\mathchar'26\mskip-9mu\lambda}_{L,U}/v_F$,
being ${\mathchar'26\mskip-9mu\lambda}_{L,U}$ the mean-field amplitude
of the Hubbard-Stratonovich fields on the
n.n.n. links.
We assume that the mean-field approximation amplitude of the
H-S fields on the n.n. links $\bar{\rho}$ is the same for both
planes. Consequently the Fermi velocity is also the same.
The only low energy
intra-layer bosonic degree of freedom left are the gauge fields of the
upper and lower planes $A_U$ and $A_L$ and the inter-layer
fields $\chi_z$.

The continuum action for the bilayer consists essentially of Eq.\ (\ref{splane})
written twice with labels $L$ and $U$ for lower and upper plane
and an inter-layer part given by the coupling between planes
\begin{eqnarray}
{\cal S}_{interlayer} & = &
\int dx_0 \int dx^2
\left\{
{\overline \psi}_L
\left(
\varphi_0\gamma_0 {\bf 1}
+\varphi_1{\bf \gamma}_1{\bf \tau}_1
+
\varphi_2{\bf \gamma}_2{\bf \tau}_2
+
\varphi_3{\bf 1}{\bf \tau}_3
\right)
\psi_U + {\bf h.c.}
\right\}
\nonumber \\
& - & {1\over g_3} \int dx_0 \int dx^2
 \left[ 
U
\left(
|\varphi_0|^2
\right)
+ U
\left(
|\varphi_1|^2
\right) 
+ U
\left(
|\varphi_2|^2
\right)
+ U
\left(
|\varphi_3|^2
\right)
\right]
\label{sinter}
\end{eqnarray}
In this expression $\psi_L$ and $\psi_U$
represent the two 
Dirac {\it flavors} $\psi^{1,2}_{L,U}$ that live on the lower 
and upper plane of the bi-layer.
The $\tau$-matrices mix Dirac flavors
inside each plane.

The intra-layer gauge fields, which represent intra-layer
phase fluctuations on n.~n.~ links have to be kept since they enter at the leading
order in the continuum limit. There are other operators, with the form of fermion mass 
terms, that have not been included
which do not contain any derivatives but they describe other types of intra-layer
ordering which compete with the CSL. To include such effects would require a theory of
the full phase diagram which is beyond the scope of this paper.

The bosonic part of the inter-layer action shown
in the second line of Eq.\ (\ref{sinter}), comes
from the corresponding bosonic terms in Eq.\ (\ref{lagrangian})
\begin{equation}
{\cal S}_b  =  - {1 \over J_3}\int dx_0 \sum_{\vec x=(e,e)}^{}
           \left\{ |\chi_z(\vec x)|^2 + |\chi_z(\vec x+{\vec e}_1)|^2
           + |\chi_z(\vec x +{\vec e}_2)|^2 + |\chi_z(\vec x +{\vec e}_1 
+{\vec e}_2)|^2\right\}
\end{equation}
where $\vec x$ is an even-even site
on the lattice at, say, the lower plane.
However, in going to the continuum limit it proves more
convenient to introduce the rotation given by
the linear combinations of the four H-S fields $\chi_z(\vec x)$ which
link corresponding {\it plaquettes} of the planes 
\begin{eqnarray}
\varphi_0(\vec x) & \approx &
{1\over 4}\left( \chi_z(\vec x) + \chi_z (\vec x + {\vec e}_1)
+\chi_z(\vec x+{\vec e}_2)+
\chi_z(\vec x+{\vec e}_1+{\vec e}_2)\right) \\
\varphi_1(\vec x) & \approx &
{1\over 4}\left( \chi_z(\vec x) - \chi_z (\vec x + {\vec e}_1) 
+ \chi_z(\vec x+{\vec e}_2) -
\chi_z(\vec x+{\vec e}_1+{\vec e}_2)\right) \\
\varphi_2(\vec x) & \approx &
{1\over 4}\left( \chi_z(\vec x) + \chi_z (\vec x + {\vec e}_1) 
- \chi_z(\vec x+{\vec e}_2) -
\chi_z(\vec x+{\vec e}_1+{\vec e}_2)\right) \\
\varphi_3(\vec x) & \approx &
{1\over 4}\left( \chi_z(\vec x) - \chi_z (\vec x + {\vec e}_1) 
- \chi_z(\vec x+{\vec e}_2)+
\chi_z(\vec x+{\vec e}_1+{\vec e}_2)\right),
\label{phies}
\end{eqnarray}
In terms of the rotated fields, and after taking the
continnum limit, the bosonic part of the
action takes the form
\begin{eqnarray}
{\cal S}_b & = & - {1 \over g_3}\int dx^3
          \left[
\varphi_0^{*}(\vec x) \varphi_0(\vec x)
              +
\varphi_1^{*}(\vec x) \varphi_1(\vec x)
              +
\varphi_2^{*}(\vec x) \varphi_2(\vec x)
              +
\varphi_3^{*}(\vec x) \varphi_3(\vec x)
\right]
\nonumber \\
& = & - N \int {dq^3 \over (2\pi)^3}
     \left[ \lambda \varphi_0^{*}(\vec q) \varphi_0(\vec q)
              + \lambda \varphi_1^{*}(\vec q) \varphi_1(\vec q)
              + \lambda \varphi_2^{*}(\vec q) \varphi_2(\vec q)
              + \lambda \varphi_3^{*}(\vec q) \varphi_3(\vec q)
\right]
\label{bosef}
\end{eqnarray}
In the second line of Eq. (\ref{bosef}), Fourier transforms
have been taken and the coupling constant $g_3$ has been
re-scaled by $\frac{1}{N}$ in order to allow a $\frac{1}{N}$-expansion
(see below). In other words $\lambda\equiv \frac{1}{{g'}_3}$, where
${g'}_3 \equiv \frac{g_3}{N}$.
The fields $\varphi_j$; $j=0,1,2,3$ also have been re-scaled
to $\varphi/v_F$. As a result,
the effective coupling constant that controls the inter-layer
fluctuations is
$ g_3 \equiv 2a {J_3\over {\overline\rho}}  = J_3 (2a)^2/v_F$
and has units of length.
Throughout this work we use dimensions such that
$[h] \ = \ [e] \ = \ [v_F] \ = \ 1$
where $h$, $e$ and $v_F$ are the Planck's constant,
the unit of charge and the Fermi velocity respectively.
We have a natural scale in our theory, which
is the lattice constant $a_0$, or the inverse
lattice constant which we shall call $\Lambda$
and characterizes the momentum cutoff.

From the free part of the action, and the fact
that we are working in $2+1$ dimensions, it is
clear that the dimension of the fermion operators
must be $ \Lambda \approx ({\rm length})^{-1}$.
The dimension of the operator
$\hat{\varphi}$ is also that of $\Lambda$.
The coupling constant $g_3$ is dimensional
with $[\lambda] \equiv \left[\frac{1}{g_3}\right] = \Lambda$.
This dimensional analysis tells us that the effective
four-fermion operator which represents the interactions between
the fermions of the two planes, is irrelevant at the weak coupling fixed
point and
that, if a phase transition exists, it should happen at some finite
value of the inter-layer coupling.  We will see  that this is indeed the
case.

Now we integrate out the fermions and obtain the effective action
\begin{equation}
{\cal S}_{eff} \equiv - i N {\rm Tr} \ln
\left[
\begin{array}{lr}
i{\raise.15ex\hbox{$/$}\kern-.70em\hbox{$D$}}_L -m_L & \hat{\varphi} \\
\hat{\varphi}^{*} & i{\raise.15ex\hbox{$/$}\kern-.70em\hbox{$D$}}_U - m_U
\end{array}
\right] + {\cal S}_b
\label{eaction}
\end{equation}
where we have defined
\begin{equation}
\hat{\varphi} \equiv \varphi_3\tau_3+\varphi_0\gamma_0+\varphi_1\gamma_1\tau_1+
\varphi_2\gamma_2\tau_2
\label{phiop}
\end{equation}
The saddle point equations are
\begin{equation}
{1\over g_3}\varphi_j^{*}(0) = -i \int {dk^3 \over (2\pi)^3} {\rm tr}  
\left[ \left(
\begin{array}{lr}
{\raise.15ex\hbox{$/$}\kern-.57em\hbox{$k$}}-
{\raise.15ex\hbox{$/$}\kern-.72em\hbox{$A$}}_L 
- m_L & \hat{\varphi}(k) \\
\hat{\varphi}^{*}(k) & {\raise.15ex\hbox{$/$}\kern-.57em\hbox{$k$}} 
-{\raise.15ex\hbox{$/$}\kern-.72em\hbox{$A$}}_U - m_U
\end{array}
\right)^{-1}\left(
\begin{array}{lr}
0 & {\delta \hat{\varphi}(k)\over \delta \varphi(0)} \\
0 & 0
\end{array}\right)\right]
\label{spe}
\end{equation}

Formally, this integral diverges linearly with the momentum cutoff
scale $\Lambda$. As in all theories of critical phenomena, we will 
absorb the singular dependence on the microscopic scale in a
renormalization of the coupling constant.
We can define a critical coupling constant $g_c$ as the value of  the
coupling constant at which the expectation values for
the fields coupling the planes first become different from zero.
Clearly the solution with $<\varphi_j^{*}> = 0$ is allowed for
any finite value of the cutoff, no matter how large. This
is the phase where
the inter-plane field is not condensed.
The non trivial solution will first occur at the value
of the coupling constant $g_j^c$ given by
\begin{equation}
{1\over g_j^c} \equiv -i \left( \frac{\delta}
{\delta \varphi_j^{*}}\right) \int {dk^3 \over (2\pi)^3}
{\rm tr} \left[\left(
\begin{array}{lr}
{\raise.15ex\hbox{$/$}\kern-.57em\hbox{$k$}} - m_L & \hat{\varphi}(k) \\
\hat{\varphi}^{*}(k) & {\raise.15ex\hbox{$/$}\kern-.57em\hbox{$k$}} - m_U
\end{array}
\right)^{-1}\left(
\begin{array}{lr}
0 & {\delta \hat{\varphi}(k)\over \delta \varphi(0)} \\
0 & 0
\end{array}
\right)\right],
\end{equation}
evaluated at the point where the $\varphi$'s vanish.

Notice that although the bare value of the coupling constants
are originally the same and equal to $g_3$, they are associated with
operators which do not scale in the same way. Their
critical values are different as well.
As an abuse of notation, from now on we are calling scalar
to the interaction channel given by the field $\varphi_3$,
frequency-vector channel to the field $\varphi_0$ and
spatial-vector channels to the ones given by $\varphi_1$
and $\varphi_2$.

Without coupling
between the planes we have a degenerate situation between
a state in which both planes have the same amount of frustration
({\it i.e.,\ } the fermion masses are the same in magnitude)
but their relative sign could be the same or opposite.
We are going to call these two states ferromagnetic (FM)
or antiferromagnetic (AFM) ordered respectively, understanding
that we refer to the relative ordering of the sign of
the chiralities.
We want to investigate how the
degeneracy between the FM and the AFM arrangement of masses
is removed. For simplicity, we give the results
for the case of $|m_L|=|m_U|= m > 0$. They may carry
any sign. We define the variable $s={\rm sign}(m_L)\;
{\rm sign}(m_U)$, which takes values $\pm 1$.
The critical values for the coupling constants are
given by
\begin{eqnarray}
\frac{1}{g_3^{c}} & = &
 i{\rm tr}\left[{\bf {\hat S}}_L(k)
\tau_3{\bf {\hat S}}_U(k)\tau_3\right]
=
\ \frac{1}{2\sqrt{\pi}}  \Lambda - \frac{1}{2\pi} m\left( 1 + s \right) \\
\frac{1}{g_0^{c}} & = &
i{\rm tr}\left[{\bf {\hat S}}_L(k)
\gamma_0{\bf {\hat S}}_U(k)\gamma_0\right]
=
\ \frac{1}{2\pi} m \left( 1 - s \right) \\
\frac{1}{g_j^{c}} & = &
i{\rm tr}\left[{\bf {\hat S}}_L(k)\gamma_j\tau_j
{\bf {\hat S}}_U(k)\gamma_j\tau_j\right]
=
\ \frac{1}{4\sqrt{\pi}} \Lambda - \frac{1}{2\pi} m \left( 1 - s \right)
\label{couplings}
\end{eqnarray}
where ${\bf {\hat S}}_a(k) 
\equiv \frac{1}{{\raise.15ex\hbox{$/$}\kern-.57em\hbox{$k$}} -m_a}$, 
with $a=L,U$ and $j=1,2$.

When the interaction between the planes is antiferromagnetic
({\it i.e.,\ } $J_3 > 0$ ) the physical coupling constants remain positive.
We are interested in the regime where $m << \Lambda$.
For the case of an AFM relative 
ordering of chiralities ({\it i.e.,\ } for $s=-1$)
we obtain
\begin{equation}
\frac{1}{g_3^c} = \frac{\Lambda}{2\sqrt{\pi}} \kern 1true cm
\frac{1}{g_0^c} = \frac{m}{\pi} \kern 1true cm
\frac{1}{g_j^c} = \frac{\Lambda}{4\sqrt{\pi}} \ - \ \frac{m}{\pi}
\end{equation}
For $m < \Lambda$, we have $g_3^c < g_j^c < g_0^c$; hence the
channel which will first undergo a transition within the
mean-field approximation, is the scalar channel,
given by the field $\varphi_3$.

On the other hand, for the case of FM relative ordering
of chiralities, we obtain
\begin{equation}
\frac{1}{g_3^c} = \frac{\Lambda}{2\sqrt{\pi}} - \frac{m}{\pi} \kern 1truecm
\frac{1}{g_0^c} = 0 \kern 1truecm
\frac{1}{g_j^c} = \frac{\Lambda}{4\sqrt{\pi}}
\end{equation}
For $m < \Lambda$, again we have $0 < g_3^c < g_j^c < g_0^c$.
Again the channel which will first undergo a transition, if any,
will be the scalar one.

In the case of ferromagnetic inter-plane coupling
({\it i.e.,\/} in the case $J_3 < 0$) 
there is no transition, 
since the critical coupling constants always
remain positive.
The exact values of the 
critical coupling constants are not universal and they depend
on the cutoff procedure that it is being used.
Our continuum approximation is not
very sensitive to these short distance features. 
However, the theory has a natural built in regulator 
since the model comes from a lattice theory. 
In other words, the qualitative feature of the
existence of critical values for the coupling constants is 
independent of the type of cutoff procedure, although their precise
value is not. The question of whether these critical values can
be physically reachable is a different issue that needs a more
detailed specification of the short distance properties of the 
model. We do not attempt to address this point here. 
We obtain the regularized saddle-point equations by
subtracting the value of $1/g_c$ on both sides
of Eq. (\ref{spe}).
\begin{equation}
\left( \frac{1}{g_3} - \frac{1}{g^{c}_j} \right) \varphi_{j}^{*} =
-i \left\{ {\rm tr} \left[ {\bf {\hat S}}_L
\frac{\delta \hat{\varphi}}{\delta \varphi_j} {\bf {\hat S}} \hat{\varphi}^{*}
\left( 1 - {\bf {\hat S}}_L\hat{\varphi}
{\bf {\hat S}}_U\hat{\varphi}^{*}\right)^{-1} \right]
- \varphi_j^{*} {\rm tr} \left[ {\bf {\hat S}}_L 
\frac{\delta \hat{\varphi}}{\delta \varphi_j}
{\bf {\hat S}}_U\frac{\delta \hat{\varphi}^{*}}
{\delta \varphi_j^{*}}\right]\right\}
\label{spefi}
\end{equation}
The simplest non trivial solution
is the one where only the scalar channel $\varphi_3$
is condensed. This channel has the
lowest critical coupling, and it will be
the first to pick a non-vanishing expectation value.
For an antiferromagnetic relative ordering of the chiralities,
which we will show it is favored in the case of antiferromagnetic
Heisenberg exchange between the planes, we find 
\begin{equation}
\left(\frac{1}{g_3}-\frac{1}{g_3^c}\right) \varphi_3^{*}  =  4 i \varphi_3^{*}
\int {dk^3 \over (2\pi)^3}
\left\{\frac{1}{
\left( k_{\mu}k^{\mu} - m^2 - |\varphi_3|^2\right) }
- \frac{1}{k_{\mu}k^{\mu}- m^2}\right\}
\label{scacond}
\end{equation}
When solving Eq.\ (\ref{scacond}) one
gets
\begin{equation}
\lambda_3  =  \left[ m - \sqrt{m^2+|\varphi_3|^2}\right]
\label{state}
\end{equation}
In Eq.\ (\ref{state}),  $\lambda_3 \equiv \frac{\pi}{g_3}-\frac{\pi}{g_3^c}$
is the distance to the critical point.
This is our equation of state.
The non-trivial solution is
\begin{equation}
|\varphi_3|^2 = \lambda_3\left(\lambda_3 -2m\right)
\label{eq5.3}
\end{equation}
It is clear from Eq. (\ref{state}) that $\lambda_3 \leq \ 0$.
When $\lambda_3 \ < \ 0$, {\it i.e.,\/} when $g_3 \ > \ g_3^c$ we find
a phase where the scalar channel field has a
nonvanishing expectation value
given by Eq. (\ref{eq5.3}).

The physics of this state is the following. The fact that $\varphi_3$
acquires an expectation value means that, on average, the inter-layer
Hubbard-Stratonovich field is different from zero. Thus, it appears that
in this state the fermions from one layer are free to go onto the other
layer. However, the corrections to this mean field picture should, among
other things, enforce the constraint of single occupancy at each site of
each layer. The only state which is compatible with the single occupancy
constraint {\it and} with inter-layer fermion hopping is a state in
which, on {\it each} link between the two layers there is a {\it spin
singlet} or {\it valence bond} state. Thus, the phase transition that we
found is a transition between two CSL states on each layer (with
antiferromagnetic ordering of the relative chiralities) and a {\it
spin gap} state with spin singlets on the inter-layer
links. A number of recent works\cite{lee,millis,scalapino} have
predicted a similar phase
transition in bilayers but between Ne{\'e}l states and spin
gap state with properties which are virtually indistinguishable from
ours.

\section{Relative ordering of chiralities}
\label{sec:order}

In this section we show that there
exists a dynamical way in which the physical system
selects a particular ordering of the chiralities in the planes.
We assume that in each plane a CSS is stabilized.
Thus, at each plane both Dirac fermion species
are coupled to the mass term with the same sign.
We assume that the mass is
the same for both fermionic flavors in each particular
plane, say $m_L$ and $m_U$ respectively.
This is consistent
with the fact that there is no explicit anisotropy present.
As in section \ref{sec:mft}, the magnitudes of the masses
are the same but their signs could be either the same or opposite.
We neglect fluctuations of the n.n. amplitude
of the H-S fields inside the planes, which can generate
a difference between the masses of the Dirac species inside
each plane, and even drive the CSS into a dimer phase
(see, for example, reference \cite{book}).

Our goal is to compute the correction to the energy of the
ground state of the bilayer system, due to the quantum fluctuations
of the fields coupling the planes. We work in the phase where no
field is condensed. Thus, the effective action derived in
section \ref{sec:mft} will describe the fluctuating part
of these bosonic fields with zero
expectation value.
The strategy is, therefore, to expand this action in powers of
(the small fluctuating part of) the fields $\varphi_0$ to $\varphi_3$
and keep up to the gaussian terms. Then, integrate the bosonic fields
out and, after re-exponentiating the expression, obtain the desired
correction to the ground-state energy density. 
This correction will contain a divergent part which is symmetric in
the sign of the masses of the fermions in different planes,
and a finite contribution
which is a function of the fermion masses of
both planes, $m_{L,U}$, with their signs.
At this point, we look for
the configuration of masses which minimizes the energy.
The case of zero mass at any plane is excluded since
we assumed beforehand that a CSS is stabilized at each plane.
This is important since these masses provide the energy gap
which is necessary for our saddle point approximation to
be stable and to allow for a semi-classical expansion.

The integration over the fermionic degrees of freedom
gives the following contribution to the effective action
(see Eq.\ (\ref{eaction})).
We have
\begin{equation}
- iN {\rm Tr} \ln \left[
\begin{array}{lr}
 i{\raise.15ex\hbox{$/$}\kern-.57em\hbox{$\partial$}}-m_L & \hat{\varphi} \\
      \hat{\varphi}^{*} & 
i{\raise.15ex\hbox{$/$}\kern-.57em\hbox{$\partial$}} - m_U
\end{array}
\right]
 = -  i N {\rm Tr} \ln\left(
\begin{array}{lr}
i{\raise.15ex\hbox{$/$}\kern-.57em\hbox{$\partial$}}-m_L & 0 \\
0 & i{\raise.15ex\hbox{$/$}\kern-.57em\hbox{$\partial$}}-m_U
\end{array}\right)
 + i N \sum_{n=1}^{\infty} {1\over n}
                   {\rm Tr} 
\left\{ \left(\hat{\bf S} \hat{\bf Q}\right)^n\right\}
\label{lnexp}
\end{equation}
Here
\[ \hat{\bf S} = 
\left(\matrix{i{\raise.15ex\hbox{$/$}\kern-.57em\hbox{$\partial$}}-m_L & 0\cr
 0 & i{\raise.15ex\hbox{$/$}\kern-.57em\hbox{$\partial$}}-m_U\cr}\right)^{-1}
\ \ \
{\rm and}
\ \ \ \
\hat{\bf Q} = \left(
\begin{array}{lr}
0 & -\hat{\varphi} \\
-\hat{\varphi}^{*} & 0
\end{array}\right)
\]

At this point it is convenient to re-scale the fluctuating fields by
${1\over \sqrt{N}}$.
Under this transformation all the terms in Eq.\ (\ref{lnexp}) that are
quadratic in the fields $\varphi$'s and ${\cal S}_b$ as expanded in
Eq.\ (\ref{bosef}) become contributions of ${\cal O}(1)$, being the
classical energy of the ground-state ({\it i.e.,\ } the classical part
of the euclidean action) of ${\cal O}(N)$.
To study the selection of the ordering of chiralities
we need to compute this ${\cal O}(1)$ correction to the ground-state
energy due to the effect of the fluctuations of the
fields coupling the planes. We first need to calculate
the one-loop contribution to the fermion determinant.
There is only one diagram to this order, which has two
external bosonic legs and two internal fermion propagators
\begin{eqnarray}
i {1\over 2} \int dx^3
\left( \hat{\bf S}\hat{\bf Q}\hat{\bf S}\hat{\bf Q}\right)
& = & i {1\over 2}\int {dk^3 \over (2\pi)^3}\int {dq^3 \over (2\pi)^3}
{\rm Tr}\left(\hat{\bf S}(k)\hat{\bf Q}(q)\hat{\bf S}(k -q)
\hat{\bf Q}(-q)\right)
\nonumber \\
& \equiv & {1\over 2}\int {dq^3 \over (2\pi)^3} 
{\cal K}^{(j)}(q)\varphi_j^{*}(q)\varphi_j(q)
\label{eq3.12}
\end{eqnarray}
As we saw before, this diagram has an ultraviolet divergence
which will be absorbed in a renormalization
of the coupling constants. So the kernels
${\cal K}^{(j)}(q)$ in Eq.\ (\ref{eq3.12}) include both the finite
part of the diagram and a contribution linearly divergent
in the integration momentum.
The computation of ${\cal K}\left( q\right)$, although rather
cumbersome is fairly straightforward.
Let us recall that we have
four channels: $\varphi_3$ can be
regarded as a scalar-like coupling to the Dirac fermions;
the other three --$\varphi_0$ to $\varphi_2$-- resemble a
gauge-field-like coupling.
This is not the
case, however, since Lorentz invariance is broken by the
presence of the $\tau$-matrices in the expression for
${\bf{\hat Q}}$.
This point is crucial. Since we do not have to
preserve Lorentz invariance when regulating the
divergent diagrams, time and space components do
not enter on equal grounds. Our theory is in
fact the continuum limit of a lattice theory.
At that level it is very clear that the only
physically sensible cutoff at hand is the
inverse lattice spacing. As a result, our
regulating procedure consists of integrating over
frequency first and then using an isotropic
gaussian cutoff for the spatial part of the
momentum. In this way we expect to recover the
qualitative features of the (finite) lattice
theory in the continuum limit.
Let us also mention that the only two spatially symmetric
combinations of the interlayer amplitudes within a
plaquette (see Eq.\ (\ref{phies})) are given by
$\varphi_3$ and $\varphi_0$.

From now on, the expressions will be given in their
Wick rotated ({\it i.e.,\/} imaginary time) form. Consequently
${q}^{\ 2}\equiv q_0^2+q_1^2+q_2^2$, where
$q_0 = -i\omega$.
We obtain (see appendix):
\begin{equation}
{\cal K}^{(3)}(q) = \frac{\Lambda}{2\sqrt{\pi}}-
{1\over 2\pi}\left\{ \frac{1}{2}\left( {q}^{\ 2} +
         m^2 \left( 1 + s \right)\right){\cal I}_0\right\}
\label{scalar}
\end{equation}
\begin{eqnarray}
{\cal K}^{(0)}(q) & = & {1\over 2\pi}
          \big\{ m \left( 1 + 2 {\kappa}_0^2 + 3 {\kappa}_0^4\right)
             + \frac{{q}^{\ 2}}{2 m}
            \left( {\kappa}_0^2 - {\kappa}_0^4\right)
\nonumber \\
    & + & \frac{1}{2}{\cal I}_0  \left[ 2 m^2 s + 4 m^2\left( \kappa_0^2
+ \frac{3}{2} \kappa_0^4\right) - \frac{{q}^{\ 2}}{2}
\left( 1 + 3 \kappa_0^4\right)\right]\big\}
\label{frequency}
\end{eqnarray}
\begin{eqnarray}
{\cal K}^{(j)}(q) & = & \frac{\Lambda}{4\sqrt{\pi}} -
{1\over 2\pi}
          \big\{ m \left( 1 + 2 {\kappa}_j^2 + 3 {\kappa}_j^4\right)
             + \frac{{q}^{\ 2}}{2 m}
           \left( {\kappa}_j^2 - {\kappa}_j^4\right)
\nonumber \\
    & + & \frac{1}{2}{\cal I}_0 \left[ 2 m^2 s + 4 m^2\left( \kappa_j^2
+ \frac{3}{2} \kappa_j^4\right) - \frac{{q}^{\ 2}}{2}
\left( 1 + 3 \kappa_j^4\right)\right]\big\}
\label{vector}
\end{eqnarray}
In Eq.\ (\ref{frequency}) and Eq.\ (\ref{vector})
${\kappa}_j^2\equiv q_j^2 / {q}^2$, \  with
$j=0,1,2$.
Notice that
the expression corresponding to the
channel given by $\varphi_0$
(loosely speaking, the frequency channel) has
an overall opposite sign to the expression for the channels given by
$\varphi_1$ and $\varphi_2$
for the finite part of the diagrams. However, the frequency channel
does not have a divergent contribution.
This sign will turn out to be quite important for the phase diagram.

On the other hand, ${\cal I}_0$ is
\begin{equation}
{\cal I}_0 = {2 \over |{q}|}\left\{
\sin^{-1}\left(
\frac{|q |}{\sqrt{ 4m^2 + {q}^{ \ 2}}}
\right)\right\}
\end{equation}

At this point, in euclidean space, we have
\begin{eqnarray}
{\cal Z}{\kern-.1truecm} & = & \int {\cal D} b e^{-E_0 - \sum_{j=0}^3
              \int {d^3 q \over (2\pi)^3}\varphi^{*}_j(q)
       \left[ \lambda -{\cal K}^{(j)}(q)  \right]\varphi_j(q)}
\nonumber \\
   & = & e^{-E_0}\prod_{j=0}^{3}\int{\cal D}\varphi^{*}_j{\cal D}\varphi_j
      \exp\left\{ - \int{d^3 q_E\over (2\pi)^3}\varphi^{*}_j(q)
      \left[ \lambda - {\cal K}_E^j(q) \right]\varphi_j(q)\right\}
\nonumber \\
   & = & \# e^{-E_0} \exp \left\{
   -\sum_{j=0}^3\int_{q} \ln\left[ \lambda - 
{\cal K}^{(j)}(q) \right]\right\}
\label{correc}
\end{eqnarray}
From Eq.\ (\ref{correc}), the correction to the energy of the ground state
due to the fluctuations of the fields $\varphi$'s is given by
\begin{eqnarray}
& & \Delta E = \int {d^3 q\over (2\pi)^3}
             \ln \left[ \lambda -
\frac{\Lambda}{2\sqrt{\pi}} +
{1\over 2\pi}\left( 2 m + \frac{1}{2}\left( {q}^{\ 2} +
         m^2 \left( 1 + s \right)\right){\cal I}_0\right) \right]
\nonumber \\
& & + \int{\kern-.1truecm} {d^3 q\over (2\pi)^3}
             \ln\big[\lambda -
{1\over 2\pi}
          \big\{ m \left( 1 + 2 {\kappa}_0^2 + 3 {\kappa}_0^4\right)
             + \frac{{q}^{\ 2}}{2 m}
           \left( {\kappa}_0^2 - {\kappa}_0^4\right)
\nonumber \\
    & + & \frac{1}{2}{\cal I}_0 \left[ 2 m^2 s + 4 m^2\left( \kappa_0^2
+ \frac{3}{2} \kappa_0^4\right) - \frac{{q}^{\ 2}}{2}
\left( 1 + 3 \kappa_0^4\right)\right]\big\}
\nonumber \\
& & +\sum_{j=1}^{2} \int  {d^3 q\over (2\pi)^3}\ln \big[\lambda -
\frac{\Lambda}{4\sqrt{\pi}} -
{1\over 2\pi}
          \big\{ m \left( 1 + 2 {\kappa}_j^2 + 3 {\kappa}_j^4\right)
             + \frac{{q}^{\ 2}}{2 m}
           \left( {\kappa}_j^2 - {\kappa}_j^4\right)
\nonumber \\
    & + & \frac{1}{2}{\cal I}_0\left[ 2 m^2 s + 4 m^2\left( \kappa_j^2
+ \frac{3}{2} \kappa_j^4\right) - \frac{{q}^{\ 2}}{2}
\left( 1 + 3 \kappa_j^4\right)\right]\big\}
\big]
\label{deltae}
\end{eqnarray}
We want to study the weak coupling regime,
which corresponds to the case of large $\lambda$ in Eq.\ (\ref{deltae}).
Moreover, this is presumably the only regime for which
Eq.\ (\ref{deltae}) is valid, since as we show later, there is
a critical value of the coupling constant at which there is
an onset of condensation for some of the interaction channels
between the planes.

By expanding in powers of ${1\over \lambda}$, to first order
we obtain that the energy correction does
not depend on the relative sign of the masses $s$ and it is completely
symmetric with respect to the exchange $m_L$ into $m_U$. This result
remains true even when the magnitude of the masses are different.
To second order we get
\begin{equation}
\Delta E^{(2)}  =  f_{symm}
 +  {1\over \lambda^2} \int {d^3 q\over (2\pi)^3}
\left( \frac{m^2}{4\pi^2}{\cal I}_0\right) s \left[ 4 m - \frac{4}{3}
\frac{m {q}^{\ 2}}{4 m^2 + {q}^{\ 2}}
+ 2 {q}^{\ 2}\left( 1 - \frac{4 m^2}{{q}^{\ 2}}\right)
{\cal I}_0\right]
\label{order2}
\end{equation}

The coefficient of $s$, where $s$ is the relative sign of the
masses ({\it i.e.,\ } of the chiralities), is a function always positive.
Thus, a minimum in the energy is obtained when $s = -1$,
which indicates that the chiralities of the planes have opposite sign.
This is the main result of this section. Recently, Gaitonde, Sajktar
and Rao\cite{gaitonde} studied  the problem of selection of the
relative chirality by means of a perturbation theory in the inter-layer
exchange coupling. They found that the ferromagnetic ordering was
selected and that this result only appeared in third order in $J_3$.
This result disagrees with ours (see Eq.\ (\ref{order2})).
It is unclear to us what is the origin of this discrepancy. The work by
Gaitonde {\it et al} relies on a rather complex lattice perturbation
theory calculation of the inter-layer correlation effects. In our work we
have evaluated the same correlation effects but within a continuum
approximation which makes the computation more transparent and easy to
check. We have used a cutoff only for the space components of the
momentum transfers in our Feynman diagrams. The form of the cutoff that
we chose closely mimics the effects of the lattice. Thus, it is unlikely
that the discrepancy could be due to different choices of cutoffs.
Similarly, the discrepancy appears at very weak inter-layer coupling
where $J_3 \ll |m|$, where $|m|$ is the magnitude of the mass of the
chiral excitations on each layer. Although it is conceivable that this
discrepancy could be due
to highly energetic processes which may be treated differently by both
cutoff procedures, this appears to be unlikely since the mass $|m|$ is
very large in this regime. Barring some numerical difficulty (which is
possible in such involved calculations), the absence of a correction
which depends
on the relative sign of the mass in the lattice calculation 
(to the same order as the one given by Eq.\ (\ref{order2})) points to
the occurrence of a special cancelation which we do not see in the
effective continuum theory. We have also checked our result with
other choices of cutoff on the space components and we have always found
the same effect. Only in one instance, when we used a relativistic form
of the cutoff, isotropic in both space and time, we found it necessary
to go to third order in $J_3$, which resembles the result 
reported by Gaitonde {\it et al}, but even in that case we found that
the {\it antiferromagnetic} ordering of chiralities is the one 
energetically favored. However, the relativistic
cutoff is certainly the one which is most unlike the lattice cutoff. In view of
this considerations, we strongly believe that our treatment is robust
and reliable.

\section{Landau-Ginzburg effective theory}
\label{sec:L-G}

We want to study the behavior of the low-energy modes for this system.
The approach we are taking here is to derive an effective
theory for the fluctuations of the $\varphi$-fields and the gauge fields.
We want to study and characterize the phase diagram at the tree level
approximation or Landau-Ginzburg approximation, and further on, investigate
the effects of the fluctuations. We showed that
there exist critical values for the coupling constants which possibly 
mark a transition between a symmetric or non-condensed phase for the
$\varphi$-fields and a phase in which at least the scalar channel 
acquires an expectation value. The Landau-Ginzburg theory to
be derived in this section will allow us to study the actual nature of
this phase transition.
We expand the fermionic determinant
in a gradient expansion for slow varying modes of the fields in which 
we are interested. 

We derive an effective action only for the scalar channel.
This particular channel is the one that first undergoes a condensation,
for de case of an antiferromagnetic ordering of the chiralities, 
since it has the lowest critical coupling constant with a positive value.
The other three channels will remain massive modes and consequently 
they can be integrated out
of the theory. This process will involve renormalization of the 
parameters of the system but it will not affect dramatically the
underlying physics.
On the contrary, the scalar channel effectively undergoes a transition
as the critical value of the coupling constant is approached and crossed.
The bosonic excitations become massless at the transition point and
we want to study the physics on both sides of this transition.
We use the following definitions 
$ A_{+}^{\mu} (x) \equiv  A_{L}^{\mu} (x) + 
 A_{U}^{\mu}(x)$ and 
$ A_{-}^{\mu} (x) \equiv A_{L}^{\mu} (x) - 
A_{U}^{\mu}(x)$
for the in phase and out of
phase gauge fields respectively. The covariant derivative is defined as
${\cal D}_{\mu} \equiv \partial_{\mu} - i \  A^{-}_{\mu}$.

The details of the calculation are described roughly in Appendix III.
The following effective action is obtained by Fourier antitransforming
the contributions of the 1-loop diagrams up to order $\frac{1}{N}$,
where $N$ is the fermion species number. This includes bubble diagrams
with up to four legs, since each of these legs represents the fluctuating
part of either a matter or a gauge field, which has been previously
re-scaled by a factor $\frac{1}{\sqrt{N}}$. The loop integration adds
a factor of $N$ coming from the number of fermions propagating in the
loop. From these diagrams we keep terms up to second order  in the
external momenta. In real space we find various terms; we get a 
contribution involving only the gauge fields which we call
${\cal S}^{(0)}_{gauge}$. This arises from the fermion loops 
corresponding to the propagation of spinon-hole pairs inside each
plane, without mixing. It contains the usual square of the field strength 
tensor and the induced Chern-Simons term. In the $ A_{(+)}$-
$ A_{(-)}$ coordinates this term is off diagonal, since the
sign of time reversal invariance is opposite between the planes,
\begin{eqnarray}
{\cal S}^{(0)}_{gauge}(x) = \frac{1}{16\pi} \int dx^{3}
\epsilon_{\mu\nu\lambda} \left( F_{(+)}^{\mu\nu}(x)
A_{(-)}^{\lambda}(x) + F_{(-)}^{\mu\nu}(x)
A_{(+)}^{\lambda}(x)\right)
\nonumber \\
- \qquad \frac{1}{64\pi|m|} \int dx^{3}
\left( F_{(+)}^{\mu\nu}(x) F^{(+)}_{\mu\nu}(x)
\ + \ F_{(-)}^{\mu\nu}(x) F^{(-)}_{\mu\nu}(x)
\right)
\label{lgauge}
\end{eqnarray}
The following term has a free part for the field $\varphi$ and
another part coupling this field to the gauge fields. 
A term coupling the gauge invariant current for the matter
field $\varphi$ to the field strength tensor of the in phase 
gauge field is also present.
\begin{eqnarray}
{\cal S}^{(1)}_{\varphi}(x) & =  & \frac{1}{4\pi |m|} \int dx^{3}
\left[\left(\partial_{\mu} + \frac{i}{\sqrt{N}} A^{-}_{\mu} (x)
\right) \varphi^{*}(x) \left( \partial^{\mu} - \frac{i}{\sqrt{N}}
A_{-}^{\mu}(x) \right)\varphi(x)\right]
\nonumber \\
&& \qquad\qquad - \ \frac{1}{32 \pi} \ \frac{1}{|m|^2} \int dx^{3}
\epsilon_{\mu\nu\lambda} F^{\mu\nu}_{(+)}(x) {\bf J}^{\lambda}_{(-)}
(x)
\label{luno}
\end{eqnarray}
In Eq.\ (\ref{luno}) we have defined the current operator for the field
$\varphi$ as 
\begin{equation}
{\bf J}^{(-)}_{\lambda}(x) 
= \ i \ \left(\varphi^{*}(x)\partial_{\lambda} 
\varphi(x) \ - \ \varphi(x)\partial_{\lambda}\varphi^{*}(x)
\right) \ + \ \frac{2}{\sqrt{N}}A^{(-)}_{\lambda} (x)|\varphi (x) |^2
\label{curr}
\end{equation}
Notice that
all the terms are manifestly gauge invariant as it should be, since
this symmetry was present before we integrated out the fermions. 
Notice also that the matter field
couples only to the out of phase or relative gauge field. This is 
consistent with the symmetry of plane exchange which remains intact
if the magnitude of the fermion mass is the same on both planes.
In other words, the original theory was invariant under the
exchange of $ A_{L}$ and $ A_U$ and the sign of the masses.
This invariance should remain at this level for our approximation to be
consistent. However, $A_{(-)}$ changes sign under this operation.
This amounts to reverse the sign of the charge, or charge conjugation 
and consequently $\varphi$ has to be conjugated. This renders the
covariant derivative term and the gauge invariant current unchanged. 
On the other hand, $F^{\mu\nu}_{(+)}$ is invariant under plane exchange.
All the other terms are even on $A_{(-)}$
and our effective action verifies the plane interchange symmetry.
Finally, from the contributions coming from the four leg diagrams 
which are of second order in the external momenta  we can derive 
the following higher derivative terms
\begin{eqnarray}
{\cal S}^{(2)}(x) & = & \frac{1}{16\pi} \ \frac{1}{|m|^3} \big\{
\int dx^3 |{\cal D}^2 \varphi(x)|^2 
 -  \frac{2}{3} \int dx^3
\left( \{ {\cal D}_{\mu},{\cal D}_{\nu} \} \varphi (x) \right)^{*}
\left( \{ {\cal D}^{\mu}, {\cal D}^{\nu} \} \varphi (x)\right) 
\nonumber \\
& & \qquad\qquad + \ \frac{1}{2} \int dx^3  F^{(-)}_{\mu\nu} (x) 
F^{\mu\nu}_{(-)}(x)|\varphi(x)|^2
+ \frac{1}{6} \int dx^3 F^{(+)}_{\mu\nu} (x) F^{\mu\nu}_{(+)}(x)
|\varphi(x)|^2
\big\}
\label{ldos}
\end{eqnarray}
We also get a self interacting term for $\varphi$ given by
\begin{equation}
{\cal S}_{self} (x) = - \ \frac{1}{N} \ \frac{1}{4\pi |m|} 
\int dx^3 |\varphi(x)|^4
\ - \ \left( \frac{1}{g} - \frac{1}{g_c} \right)
\int dx^3 |\varphi(x)|^2
\label{self}
\end{equation}
In Eq. \ (\ref{self}) above, we use the definition for $g_c$ introduced in
section \ref{spe}, {\it i.e.,\ } $\frac{1}{g_c} = \frac{\wedge}{2\sqrt{\pi}}
$.
In order to re-write this effective action in a simplified way we
introduce some field re-scaling and define the following coupling costants,
\begin{eqnarray}
\varphi(x) \equiv \sqrt{4\pi|m|} \phi(x);  \quad
A^{\mu}_{(-)} (x) & \equiv & \sqrt{N}  A^{\mu}_{(-)}; \quad
m_0^2 \equiv 4\pi|m| \left(\frac{1}{g} - \frac{1}{g_c}\right); \quad
\lambda \equiv \frac{4\pi|m|}{N}; 
\nonumber \\
\theta  \equiv  \frac{N}{2\pi}; \quad
{\bf e}^2 & \equiv & \frac{16\pi|m|}{N}; \quad 
G_A \equiv \frac{1}{8|m|}; \quad \
{\rm and} \quad \
{\bar G} \equiv \frac{1}{4m^2}.
\nonumber
\end{eqnarray}
By plugging all of these in,  we obtain
\begin{equation}
{\cal L}_{gauge} = \frac{\theta}{8} 
\left( {\tilde F}^{(+)}_{\lambda}(x)
A_{(-)}^{\lambda}(x) + {\tilde F}^{(-)}_{\lambda}(x)
A_{(+)}^{\lambda}(x)\right) \
- \ \frac{1}{4{\bf e}^2} 
\left( F_{(+)}^2 (x) 
\ + \ F_{(-)}^2 (x)
\right)
\end{equation}
where ${\tilde F}_{\lambda} \equiv \epsilon_{\mu\nu\lambda} F^{\mu\nu}$
is the dual of the field strength tensor. 
We also get
a Lagrangian density for the field $\phi$ given by 
\begin{equation}
{\cal L}_{\phi} = \partial^{\mu} \phi^{*}\partial_{\mu} \phi
- m_0^2 |\phi|^2 
- \lambda |\phi|^4 + {\cal L}_{I}
\label{lfree}
\end{equation}
where
\begin{eqnarray}
{\cal L}_{I} & \equiv & 
 i \left( \phi^{*}(x)\partial_{\mu}\phi(x) - 
\phi(x)\partial_{\mu}\phi^{*}(x)\right)
\left[
A_{(-)}^{\mu}(x) - G_A {\tilde F}^{\mu}_{(+)}(x)\right]
\nonumber \\
& + & |\phi(x)|^2 
\left[ A^2_{(-)}(x) - 2 G_A {\tilde F}_{(+)}^{\mu}(x) A^{(-)}_{\mu}(x)
+ {\bar G} \left( \frac{1}{2} \ F_{(-)}^{2}(x) + \frac{1}{6} \ F_{(+)}^{2}
(x)\right)\right]
\label{intl} 
\end{eqnarray}
In Eq. \ (\ref{intl}) we dropped the higher derivative terms which
appear in Eq.\ (\ref{ldos}), except for the antisymmetric parts which
involve renormalizations of the (different) effective charges for the
in phase and out of phase gauge fields.

\section{Symmetric Phase}
\label{sec:symm}

In this section we want to study the physics of the regime in which there 
is no condensation of the field $\phi$ for the effective theory derived
in the previous section, {\it i.e.,\ } where $\phi$ has a vanishing vacuum
expectation value. This phase consists of the bilayer system with 
relatively opposite broken time
reversal invariance between both planes, but the difference with the case
of decoupled planes is that they are now linked through the fluctuations
of the field $\phi$. This field represent a massive boson like mode with
mass given by $m_0$ defined in the previous section by
\begin{equation}
m_0^2 \ = \ \frac{1}{g} - \frac{1}{g_c}  = \frac{v_F}{(2a)^2 J_3} -
\frac{\wedge}{2\sqrt{\pi}} = \frac{\rho}{2a J_3} - \frac{\wedge}{2\sqrt{\pi}}
\label{mass}
\end{equation}
The magnitude of this mass measures the distance to the critical point.
In this phase we are on the side of the transition in which $m_0^2 \ > \ 0$.
It clearly corresponds to a weakly interplane coupling regime,
{\it i.e.,\ } the limit of small $J_3$. The $\phi$-field can be integrated
out to get an effective action for the gauge fields only. 
However, for our approximations to be consistent we need to assume that 
$m_0^2$ is much smaller that the fermion mass $m$. In other words, our 
results are valid on a window not to close to the phase transition (where
$\phi$ becomes massless as $g \rightarrow g_c$ and $m_0 \rightarrow 0$)
but also not too far from the transition so that the mass of the collective
mode represented by $\phi$ never becomes comparable to the fermion mass.

We are going 
to show that there is no renormalization of coefficient of 
the Chern-Simons terms that
had been induced by the fermionic fluctuations on the planes, arising from
the fluctuations of $\phi$, at least to order $\frac{1}{N}$. There are
``charge" renormalizations in the sense that the coefficients of the
field strength tensor for both $ A_{(+)}$ and $A_{(-)}$ get
renormalized. Furthermore, we will show that the spectrum of low energy
excitations in this phase has two massive photons, whose masses do not
violate the gauge symmetry but they break parity and time reversal 
invariance, and are very effective in taming the fluctuations of the
gauge fields. In a sense we still have pretty much the same physical 
picture corresponding to two decoupled chiral spin liquid with opposite
relative breaking of time reversal invariance. Consequently we will
still have deconfined spinon as the elementary excitations of the system.
The issue of the statistic of the quasiparticles is a little more involved
as we discuss below.

Starting from the effective action derived in the previous section we 
can integrate out perturbatively the field $\phi$. This gives a 
result valid within the region of applicability of the gradient expansion.

The integration over $\phi$ gives the effective action
(for small 
${A}_{\mu}$)
We have,
\begin{eqnarray}
& & \int {\cal D}\phi {\cal D}\phi^{*} \exp{i{\cal S} \left(\phi, {A}_{\mu}
\right)}
\nonumber \\
& & \quad = 
\exp{ \left( i  {\cal S}_{gauge} ( A_{\mu})\right)}
\int {\cal D}\phi {\cal D}\phi^{*}
\exp{i \int dx^3 {\cal L}_{\phi} } \left( 1 + i{\cal L}_I - \frac{1}{2}
\left( {\cal L}_I\right)^2\right)
\nonumber \\
& & \quad =  Z_0 \exp{ \left(i {\cal S}_{gauge} \right)}
\Big\{ 1   +  i\Big< \ i \left( \phi^{*}\partial_{\mu}\phi - 
\phi\partial_{\mu}\phi^{*}\right)\Big>
\left[
A_{(-)}^{\mu} - G_A {\tilde F}^{\mu}_{(+)}\right]
\nonumber \\
& & \quad \quad + \ i \ \Big< |\phi|^2 \Big> 
 \left[ A^2_{(-)} - 2 G_A {\tilde F}_{(+)}^{\mu} A^{(-)}_{\mu}
+ {\bar G} \left( \frac{1}{2} \ F_{(-)}^{2} + \frac{1}{6} \ F_{(+)}^{2}
\right)\right]
\nonumber \\
& & \quad\quad - \ \frac{1}{2}
\Big< i^2 \left( \phi^{*}\partial_{\mu}\phi  -  
\phi\partial_{\mu}\phi^{*}\right)
\left( \phi^{*}\partial_{\nu}\phi  -  
\phi\partial_{\nu}\phi^{*}\right) \Big>
\left[
A_{(-)}^{\mu} - G_A {\tilde F}^{\mu}_{(+)}\right]
\left[
A_{(-)}^{\nu} - G_A {\tilde F}^{\nu}_{(+)}\right]
\Big\}
\label{cumulants}  
\end{eqnarray}

The cumulant coefficients can be computed in the usual way\cite{book} to find
\begin{equation}
i \ \Big< |\phi(x)|^2\Big>  =    
-  \int dq^3 \frac{1}{q^2-m_0^2} = - \ \#_{\it sing} \ - \ \frac{i}{4\pi} |m|
\label{phicuad}
\end{equation}
and
\begin{equation}
-  \frac{1}{2} \ \Big<\left( i \left( \phi^{*}\partial_{\lambda}\phi - 
\phi\partial_{\lambda}\phi^{*}\right)\right)
\left( i \left( \phi^{*}\partial_{\eta}\phi - 
\phi\partial_{\eta}\phi^{*}\right)\right)\Big>
= \left(\#_{\it sing} + \frac{i}{4\pi} |m|\right) g_{\lambda \eta}
\label{currcuad}
\end{equation}

Both integrals in Eq.\ (\ref{phicuad}) and Eq.\ (\ref{currcuad})
have a linear ultraviolet divergence and need to be regularized.
One can use any of the usual regulators, for example Pauli-Villars
or minimal subtraction (which is equivalent to an  
analytical continuation of the negative argument gamma function) 
However the finite part of both integrals 
after we treated them with the same regulating scheme is exactly the
same but with opposite sign. 
This should be the case since it is required to preserve 
gauge invariance. In other words, we cannot generate a $A_\mu A^\mu$
term in the symmetric phase because such a term would manifestly break
gauge invariance and we know this is not the case. 
Therefore, the
term $A^\lambda A_\lambda \times \Big< |\phi|^2 \Big>$ in the
r.h.s. of Eq.\ (\ref{cumulants}) should cancel 
exactly (and it does) the term 
$A^\lambda \ A^\eta \times \Big<i^2 \left( \phi^{*}\partial_{\lambda}
\phi - 
\phi\partial_{\lambda}\phi^{*}\right)
 \left( \phi^{*}\partial_{\eta}\phi - 
\phi\partial_{\eta}\phi^{*}\right)
\Big>
$.
Notice that for the same token the term which could have given a
renormalization of
the cross Chern-Simons terms get canceled. In a sense, it is also gauge 
invariance which prevents the cross Chern-Simons terms to get renormalized.

A minimal subtraction procedure will consist in the complete removal of
the singular part. In fact, any cutoff procedure which preserves gauge
invariance would work as well. It can be shown that our regularization
prescription is entirely equivalent to the introduction of a gaussian
spherical cutoff in the imaginary frequency (or euclidean) reciprocal
phase space. A term of the form $\exp{\left[ -(\frac{\pi}{\wedge^2})
(q_E^2 + m_0^2)\right]}$ does the job for us.
One should be aware however, that this cutoff is not exactly the same
used in sections \ref{sec:mft} and \ref{sec:order} since there the
cutoff was gaussian isotropic on the spatial components of the momentum
but the frequency range was unbounded. Here that cutoff procedure would
not work because it breaks gauge invariance. In sections
\ref{sec:mft} and \ref{sec:order} gauge invariance was not at stake
and we were trying to implement a regularization that resembles closely
what happens on a lattice. It should also be noticed that, although 
aparently the same field $\varphi$ is involved in both cases, we
were dealing before with ultraviolet divergences of a fermion loop
integral, while here the field propagating is the bosonic field
$\varphi$ itself. In other words, we were dealing in the previous 
sections with the self-energy of the field $\varphi$ while here we
are dealing with the self-energy of the photon or the gauge fields.
Finally we do get renormalizations for the $F_{(+)}^2$ and $F_{(-)}^2$
terms.

After this procedure is applied, we are left with the regularized
(finite) form of Eq.\ (\ref{cumulants})
\begin{eqnarray}
Z_{reg} & = &
\exp\left(i{\cal S}_{gauge}\right)
\Big\{ 1 + \frac{i}{4\pi} \ |m| \ \left[ \frac{{\bar G}}{2} F_{(-)}^2
+ \frac{{\bar G}}{6} F_{(+)}^2
- 2 \ G_A^2 \ F_{(+)}^2 \right]\Big\}
\nonumber \\
& \approx & \exp\left( i \ \int dx^3 {\cal L}_{eff} \right)
\label{effz}
\end{eqnarray}
where 
\begin{eqnarray}
{\cal L}_{eff} & = &
\frac{\theta}{8} 
\left( {\tilde F}^{(+)}_{\lambda}
A_{(-)}^{\lambda} + {\tilde F}^{(-)}_{\lambda}
A_{(+)}^{\lambda}\right) 
\nonumber \\
& - &  \left( \frac{1}{4{\bf e}^2} - \frac{1}{4\pi} \ |m| \ \frac{{\bar G}}{2}
\right) \ F_{(-)}^2 
\ - \ 
\left( \frac{1}{4{\bf e}^2} - \frac{1}{4\pi} \ |m| \ \left[\frac{{\bar G}}{6}
- 2 G_A^2\right]
\right)
F_{(+)}^2 
\label{lreg}
\end{eqnarray} 
In Eq.\ (\ref{lreg}) we used that ${\tilde F}_{\lambda}^{(+)}
{\tilde F}^{\lambda}_{(+)} = 2 F_{(+)}^2$.

We now explore the energy momentum dispersion relation.
The low energy collective modes are fluctuations of the gauge fields.
We will show that there exists a photon-like mode but it is massive.
This is of great importance for the survival of spinons in the energy
spectrum (see for example ref. \cite{book}).
The regularized (finite) theory has the form
\begin{equation}
{\cal S}_{eff} \left( A_\mu\right)
=
\int dx^3 \Big[ - \ c_{-} F_{(-)}^2 - \ c_{+} F_{(+)}^2 
+ \frac{\theta}{8} \epsilon_{\mu\nu\lambda} 
\left( F^{\mu\nu}_{(+)} A^{\mu}_{(-)}
+  F^{\mu\nu}_{(-)} A^{\mu}_{(+)} \right) \Big]
\end{equation}
where
$c_{-} = \frac{1}{64\pi|m|} \ (N-2)$; 
\ and \ $c_{+} = \frac{1}{64\pi|m|} \ (N - \frac{1}{6})$.
In momentum space we have
\begin{eqnarray}
& & {\cal S}_{eff} \left(A_\mu\right)
= 
- \int_{p} \left(A^{\mu}_{(-)}(p)  A^{\nu}_{(+)}(p) \right)
\left[
\begin{array}{lr}
2c_{-}(p^2 g_{\mu\nu} - p_\mu p_\nu) & i\frac{\theta}{4} 
\epsilon_{\mu\lambda\nu} p^{\lambda}\cr
i\frac{\theta}{4}\epsilon_{\mu\lambda\nu} p^{\lambda} & 
2c_{+} (p^2 g_{\mu\nu} -
p_\mu p_\nu) \cr
\end{array}
\right]
\left(
\begin{array}{lr}
A^{\mu}_{(-)}(-p) \cr 
A^{\nu}_{(+)}(-p) \cr
\end{array}
\right)
\nonumber \\
&& \equiv 
\int_{p} A^{\mu}_{a}(p) \Big[ c_0 (p^2 g_{\mu\nu} - p_\mu p_\nu)
 {\bf I}^{a \ b} + c_3 (p^2 g_{\mu\nu} - p_\mu p_\nu)
 {\bf T}_{3}^{a \ b} + \kappa_0 \epsilon_{\mu\lambda\nu} p^\lambda
 {\bf T}_{1}^{a \ b} \Big] A^{\nu}_{b}
\label{symminvprop}
\end{eqnarray}
with $a, b = (-), (+)$ and $\mu,\nu$ the usual Lorentz indices,
$c_0 = - (c_{-} + c_{+})$, \  $c_3 = - (c_{-} - c_{+})$ \ and \
$\kappa_0 = -i \ \frac{\theta}{4}$. 

This is a bilinear form in $A^{\mu}_a$ and the propagator for the gauge
fields is just the inverse of the matrix shown in Eq.\ (\ref{symminvprop}).
However, this matrix is singular unless we fix a gauge. This is so because
the gauge field propagator $\Big< A^\mu_a \ A^\nu_b\Big>$ is not a gauge
invariant operator and does not have a physical meaning unless we are
working in a particular gauge. We need to add gauge fixing terms in order
get the propagator. We may add for example, 
$-  \frac{1}{\alpha} \left( \partial_\mu A^\mu_{(-)}\right)^2$ 
and $-  \frac{1}{\beta} \left(
\partial_\mu A^\mu_{(+)}\right)^2$, which in momentum space take the
simple form $- \frac{1}{\alpha} p_\mu p_\nu$ and 
$- \frac{1}{\beta} p_\mu p_\nu$.
The three operators ${\hat P}_{\mu\nu}\equiv p_\mu p_\nu$, ${\hat G}_{\mu\nu}
\equiv p^2 
g_{\mu\nu}$ and ${\hat K}_{\mu\nu} \equiv \epsilon_{\mu\lambda\nu} p^\lambda$
satisfy a closed algebra, and now the matrix can be inverted.
After some lengthy though fairly straightforward algebra one gets
\begin{eqnarray}
D(c_{-}, c_{+}, \theta, \alpha,\beta)
=
\left(
\begin{array}{lr}
{\hat D}_{\mu\nu}^{(-)(-)} 
&
{\hat D}_{\mu\nu}^{(-)(+)}
\cr
{\hat D}_{\mu\nu}^{(+)(-)}
&
{\hat D}_{\mu\nu}^{(+)(+)}
\end{array}
\right)
\end{eqnarray}
given by
\begin{equation}
{\hat D}_{\mu\nu}^{(-)(-)}
=
-\frac{1}{2 \ c_{-}} \left( g_{\mu\nu} - \frac{p_\mu \ p_\nu}{p^2}\right)
\ \frac{1}{p^2 - M_{ph}^2} \ - \ \alpha \ \frac{p_\mu \ p_\nu}{p^2}
\label{d--}
\end{equation}
\begin{equation}
{\hat D}_{\mu\nu}^{(+)(+)}
=
-\frac{1}{2 \ c_{+}} \left( g_{\mu\nu} - \frac{p_\mu \ p_\nu}{p^2}\right)
\ \frac{1}{p^2 - M_{ph}^2} \ - \ \beta \ \frac{p_\mu \ p_\nu}{p^2}
\label{d++}
\end{equation}
\begin{equation}
{\hat D}_{\mu\nu}^{(-)(+)}
=
{\hat D}_{\mu\nu}^{(+)(-)}
=
i \ \frac{\theta}{4} \ \frac{1}{4 \ c_{-}c_{+}} \ \frac{1}{p^2}
\epsilon_{\mu\lambda\nu} \ p^{\lambda} \ \frac{1}{p^2 - M_{ph}^2}
\label{d-+}
\end{equation}
where we defined 
\begin{equation}
M_{ph} = \sqrt{\frac{\theta^2}{64 \ c_{-}c_{+}}}
=
\frac{8\pi|m|\theta}{\sqrt{(N-2)(N- 1/6)}}
=\frac{4 \ |m|}{\sqrt{(1-\frac{2}{N})(1-\frac{1}{6N})}}
\end{equation}
as the ``photon" mass~\cite{jackiw}. 
To leading order in $\frac{1}{N}$ we can rotate back to the  $A_L$,  $A_U$ 
coordinates to get (in th Lorentz gauge $\alpha \ = \ \beta \ = \ 0$)
\begin{equation}
{\hat D}_{LL} 
= 64 \pi \frac{1}{N}  |m|  \frac{1}{p^2 - M_{ph}^2} 
\left[ 
\left( 
g_{\mu\nu} - \frac{p_\mu \ p_\nu}{p^2}
\right) + 
4i \ |m| \epsilon_{\mu\lambda\nu} 
\frac{p^{\lambda}}{p^2}
\right]
\end{equation}
and
\begin{eqnarray}
{\hat D}_{UU} = 
64 \pi \ \frac{1}{N} \ |m| \ \frac{1}{p^2 - M_{ph}^2} 
\left[ 
\left( 
g_{\mu\nu} - \frac{p_\mu  p_\nu}{p^2}
\right) - 
4i  |m| \epsilon_{\mu\lambda\nu} 
\frac{p^{\lambda}}{p^2}
\right]
\end{eqnarray}
To next order in $\frac{1}{N}$ corrections we find additional
off diagonal symmetric mixing terms.

\section{Broken symmetry phase}
\label{sec:higgs}

In this section we want to study the phase where the matter field
$\phi$ condenses. Let us assume that we went through the critical
point into the phase where $m_0^2$ in Eq.\ (\ref{mass}) becomes negative.
From Eq.\ (\ref{lfree}) we now have another possible solution with
finite $\Big< \phi\Big>$, which actually minimizes the energy. 
This is the usual non-trivial solution for a double-well effective
potential of a $\phi^4$ theory. When $m_0^2$ becomes negative, the
solution $\phi = 0$ now becomes a local maximum instead of a minimun.
The value of the new local minimun can be obtained by minimizing
Eq.\ (\ref{lfree}) to be ${\bar \phi}_0^2 = - \ \frac{m_0^2}{2\lambda}$,
where we are using the definitions given in section \ref{sec:L-G}.
If we plug in this constant value of $\phi_0$, Eq.\ (\ref{intl})
becomes
\begin{equation}
{\cal L}_I = \phi_0^2 
\left[ A^2_{(-)}(x) - 2 G_A {\tilde F}_{(+)}^{\mu}(x) A^{(-)}_{\mu}(x)
+ {\bar G} \left( \frac{1}{2} \ F_{(-)}^{2}(x) + \frac{1}{6} \ F_{(+)}^{2}
(x)\right)\right]
\end{equation}
As in the symmetric case we have
\begin{equation}
{\cal L}_{gauge} = \frac{\theta}{8} 
\left( {\tilde F}^{(+)}_{\lambda}(x)
A_{(-)}^{\lambda}(x) + {\tilde F}^{(-)}_{\lambda}(x)
A_{(+)}^{\lambda}(x)\right) \
- \ \frac{1}{4{\bf e}^2} 
\left( F_{(+)}^2 (x) 
\ + \ F_{(-)}^2 (x)
\right)
\end{equation}
Now our effective action for the gauge fields reads
\begin{equation}
{\cal S}_{eff} \left( A_\mu\right)
=
\int dx^3 
\Big[ 
\phi_0^2 \ A_{(-)}^2 + 
\left(
\frac{\theta}{8}-\phi_0^2 G_A
\right)
 \epsilon_{\mu\nu\lambda} 
\left( F^{\mu\nu}_{(+)} A^{\mu}_{(-)}
+  F^{\mu\nu}_{(-)} A^{\mu}_{(+)} \right) \Big]
-  c_{-} F_{(-)}^2 -  c_{+} F_{(+)}^2 
\label{seffcond}
\end{equation}
where the new coefficients are
\begin{equation}
\phi_0^2 = - \ \frac{m_0^2}{2 \lambda} = - \ \frac{N}{8\pi|m|} \ m_0^2;
\end{equation}
where now $m_0^2 \ < \ 0$
\begin{equation}
c_{-} = \frac{1}{4{\bf e}^2} - \phi_0^2 \frac{{\bar G}}{2}
=
\frac{N}{64\pi|m|} \left( 1 +  \frac{m_0^2}{m^2}\right); 
\end{equation}
\begin{equation}
c_{+} = \frac{1}{4{\bf e}^2} - \phi_0^2 \frac{{\bar G}}{6}
=
\frac{N}{64\pi|m|} \left( 1 + \frac{1}{3} \frac{m_0^2}{m^2}\right); 
\end{equation}
and we define $\kappa_0$ to be
\begin{equation}
\kappa_0 = - 2i \left( \frac{\theta}{8} - \phi_0^2 G_A\right)
=
- i \ \frac{\theta}{4}\left( 1 + \frac{1}{4} \ \frac{m_0^2}{m^2} \right)
=
- i \ \frac{N}{8\pi}\left( 1 + \frac{1}{4} \ \frac{m_0^2}{m^2} \right)
\end{equation}
We still need a gauge fixing term for $A_{(+)}$. 
In this way we recover the structure of Eq.\ (\ref{symminvprop})
with minor changes.
The photon mass has changed to
\begin{equation}
M_{ph} = \sqrt{ 16 m^2 \left(1 - \frac{4}{3} \frac{m_0^2}{m^2}\right) }
\approx
\ 4 \ |m| \ \left( 1 - \frac{2}{3} \ \frac{m_0^2}{m^2}\right)
\label{phmass} 
\end{equation}
The propagator for the out of phase field $A_{(-)}$ still corresponds
to a massive field, which also has a longitudinal component
\begin{equation}
{\hat D}_{\mu\nu}^{(-)(-)} =
- \frac{1}{2 \ c_{+}}\left( g_{\mu\nu} - \frac{p_\mu \ p_\nu}{p^2}\right)
\
\frac{1}{p^2 - M_{ph}^2} \ - \ \frac{p_\mu \ p_\nu}{\phi_0^2}
\label{d--cond}
\end{equation}
\begin{eqnarray}
{\hat D}_{\mu\nu}^{(+)(+)}
& = &
-
\frac{1}{2 \ c_{+}} 
\left(
1 - \frac{\phi_0^2}{2 c_{-}} \frac{1}{M_{ph}^2}
\right)
\left( g_{\mu\nu} - \frac{p_\mu \ p_\nu}{p^2} \right)
\ \frac{1}{p^2 - M_{ph}^2} 
\nonumber \\
& & \qquad \qquad  -
\ \frac{\phi_0^2}{4 c_{-} c_{+} M_{ph}^2} \ 
\left( g_{\mu\nu} - \frac{p_\mu \ p_\nu}{p^2}\right) \
\frac{1}{p^2} 
\ - \ \beta \ \frac{p_\mu \ p_\nu}{p^2}
\label{d++cond}
\end{eqnarray}
\begin{equation}
{\hat D}_{\mu\nu}^{(-)(+)}
= 
{\hat D}_{\mu\nu}^{(+)(-)}
=
i \ \frac{\theta}{4} 
\left( 1 + \frac{m_0^2}{4m^2}\right)
\frac{1}{4 \ c_{-}c_{+}} \ \frac{1}{p^2}
\epsilon_{\mu\lambda\nu} \ p^{\lambda} \ \frac{1}{p^2 - M_{ph}^2}
\label{d-+cond}
\end{equation}
Eqs.~(\ref{d--cond}),(\ref{d++cond}),(\ref{d-+cond}) 
give  the propagators of the gauge fields
in the condensed phase. By assumption $m_0^2$ is a small parameter, 
since our approximation is valid for the vicinity of the phase transition where 
$m_0^2$ is small 
(in units of the fermion mass) measures the distance to the critical point.
Notice that, in this phase, the expansion in powers of $\frac{1}{N}$
has become an expansion in powers of this new parameter.
This is consistent with our approximation because a gradient expansion
amounts to an expansion in powers of an inverse (large) length 
scale, which in our case is set by the fermion mass or, in other
words, by the spinon gap of the decoupled system.
On the other hand, the ``photon" mass is fairly large in this phase.

We conclude this section with a qualitative 
description of the excitation spectrum in
the Broken Symmetry Phase. Let us look first at the gauge excitations.
In looking at Eq.\ (\ref{d++cond}) 
we notice the existence of a {\it massless} gauge
mode in the spectrum. 
The existence of this massless mode implies that any excitation
which couples to the $A_{(+)}$ component of the gauge field 
experiences an effective
{\it long range} (logarithmic) force mediated by the massless mode. 
In particular, the
{\it spinon} excitations of the individual planes 
(which are semions in the unbroken
phase) become permanently confined by the massless gauge fields. 
Recall that the logarithmic force is actually replaced  by a confinig potential 
due to the strong fluctuations of the gauge fields 
dominated by monopole-like configurations~\cite{book,wiegmann,polyakov}.
In a sense, this makes fractional statistics unobservable since the 
quasiparticles which were able to bear it are no longer present in the spectrum.
This spectrum is consistent with the fact that the statistical parameter 
$\theta$ is not well defined anymore in this phase (it is no longer a
topological number) since, as can be seen in Eq.\ (\ref{seffcond}) it
is now modified by a term proportional to the magnitude of the order
parameter $\phi_0$.
Also, in the broken symmetry phase, the time reversal breaking mass 
coming from the coefficient
of the Chern-Simons term is no longer effective in controling the 
fluctuations of this particular mode. Actually, in this phase, the 
Higgs mechanism that takes place conspires to give a mass to the gauge field
$A_{(-)}$, breaking spontaneously its gauge symmetry, 
while leaving the in phase
field $A_{(+)}$ untouched. 
In some sense the breaking of the phase symmetry enables
the in-phase gauge field to become massless.
 
This phenomenon is in striking contrast
with the conventional Higgs-Anderson mechanism in which a spontaneously broken
symmetry renders a gauge field massive.
The remaining out-of-phase component is massive and its mass is huge (see
Eq.\ (\ref{phmass})), {\it i.~e.\/} of the order of the fermion mass. 
This huge mass
supresses the fluctuations of the field $A_{(-)}$ and, in this manner,
it restores the broken time reversal 
invariance that was present in the decoupled bi-layer system. 
In particular, this
spectrum implies that the only allowed fluctuations 
of the bilayer system are such
that the chiralities of the planes become rigidly locked {\it locally}. 
Only in-phase,
long wavelength fluctuations of the chiralities are allowed. 
Since the two chiralities
have opposite sign, we conclude that, in this phase, there is a {\it local
cancellation} of the chiralities of the planes. Hence, chiral fluctuations are
eliminated from the physical spectrum.
Recall ~\cite{fidel} that if a Chern-Simons 
term were to be present, the monopole
configurations would be suppressed and fractional statistics would become 
observable. This is precisely what happen in the symmetric phase.

The spectrum that results from our analysis of the phase with broken symmetry is
strikingly similar to the spectrum of the bilayer system in  the singlet phase
discussed by Sandvik and Scalapino~\cite{scalapino} and Millis and
Monien~\cite{millis}. 
In fact, we believe that the two phases are the same
phase and that the broken symmetry phase is a phase with spin
singlets connecting the two layers.

\section{Conclusions}
\label{sec:conc}

In this paper we have reconsidered the problem of the selection of the
relative sign of the chiralities of two planes with Chiral Spin Liquid
states coupled via an exchange interaction. We found that the exchange
coupling selects the {\it antiferromagnetic} ordering of chiralities
and, thus, that $T$ and $P$ are not broken in bilayers.  This result
holds for both signs of the inter-layer exchange constant $J_3$.
Hence, even if each plane has a net chirality, the bilayer
system does not. Such a system will not give rise to any unusual optical
activity in light scattering experiments.
 We  determined the phase diagram of the
bilayer system and found a phase transition to a valence bond (or spin
gap) state. 
Our analysis reveals the presence of an unusual ``anti-Higgs-Anderson"
mechanism which is responsible for wiping out all trace of broken time reversal
invariance in the valence-bond state.
In a separate publication we will report on results on the quantum
numbers of the excitations and on the form of the wave function for the
bilayer system.

\section{Acknowledgements}

We are grateful to A.~Rojo for discussions on his work on the ordering
of chiralities. This work was supported in part by NSF grants No.~
DMR91-22385 at the
Department of Physics of the University of Illinois at Urbana-Champaign,
and DMR89-20538/24 at the Materials Research Laboratory of the
University of Illinois.

\newpage

\appendix

\section{Computation of the bubble diagrams}
\label{sec:B}

In this section we go through the computation of the
correction to the energy of the ground state.
The expressions given by
Eq.\ (\ref{scalar}) to Eq.\ (\ref{vector}) are obtained
from a one-loop diagram. 
In momentum space,
\begin{equation}
{\cal K}(\vec q) 
=  \  i \int {dk^3 \over (2\pi)^3} 
{{\bf T}_r\left[\left({\raise.15ex\hbox{$/$}\kern-.57em\hbox{$k$}}+m_L\right)
{\hat{\varphi}}(\vec q)\left({\raise.15ex\hbox{$/$}\kern-.57em\hbox{$k$}}-
{\raise.15ex\hbox{$/$}\kern-.57em\hbox{$q$}}+m_U\right)
     {\hat{\varphi}}^{*}(-\vec q)\right] \over 
\left( {\raise.15ex\hbox{$/$}\kern-.57em\hbox{$k$}}
{\raise.15ex\hbox{$/$}\kern-.57em\hbox{$k$}}-m_L^2\right)
       \left(\left({\raise.15ex\hbox{$/$}\kern-.57em\hbox{$k$}}-
{\raise.15ex\hbox{$/$}\kern-.57em\hbox{$q$}}\right)
\left({\raise.15ex\hbox{$/$}\kern-.57em\hbox{$k$}}-
{\raise.15ex\hbox{$/$}\kern-.57em\hbox{$q$}}\right)-m_U^2\right)}
\label{kernel}
\end{equation}
with the definitions of section \ref{sec:mft}.
In computing the expression given above the following identities
involving $2+1$-dimensional Dirac $\gamma$-matrices will be important
\begin{eqnarray}
\gamma_{a}\gamma_{b}  =  g_{ab}+ i \epsilon_{abc}\gamma^{c};
\qquad
{\rm tr}\left(\gamma_{a}\gamma_{b}\right)  & = & 2 g_{ab};
\qquad
{\rm tr}\left(\gamma_{a}\gamma_{b}\gamma_{c}\right)  
= 2 i \epsilon_{abc}
\nonumber \\
{\rm tr}\left(\gamma_{a}\gamma_{b}\gamma_{c}\gamma_{d}\right)  
& = &
2\left( g_{ab}g_{cd}+g_{ad}g_{bc}-g_{ac}g_{bd}
 \right)
\end{eqnarray}
The integral in Eq.\ (\ref{kernel}) needs to be regularized, {\it i.e.,\/} 
we need to cutoff the unphysical ultraviolet divergence due to the
integration over momentum $\vec k$.
There is
a natural cutoff in the original theory, which is the lattice spacing.
However, in going to the continuum approximation we encounter the
usual field theory divergences.

There are several methods for regulating this type of integrals. The
important point is that they should preserve the physical symmetries
involved in the problem. For the case of a gauge field there are
well established procedures such as the Pauli-Villars or the
dimensional regularization methods. It can be shown that they
preserve transversality ({\it i.e.,\ } gauge invariance).
In the case of the $\varphi$-fields, we do not have such a
symmetry to preserve. In fact, not even Lorentz invariance is
preserved. We have a length scale given by the lattice
spacing, which in turn provides the
momentum cutoff $\Lambda$ that was mentioned in Sec.( \ref{sec:mft}).
On the lattice there is no cutoff for the frequency integral.
Thus, our regulating procedure is as follows. First we perform a
subtraction at the level of the integrands.
This is equivalent to write the kernels in Eq.\ (\ref{kernel})
in the following way
\begin{equation}
{\cal K}_j(\vec q) \equiv {\cal K}_j(0) + \left[ {\cal K}_j(\vec q) 
- {\cal K}_j(0) \right]
\end{equation}
In this expression only the first term has an ultraviolet
divergence. The term between brackets is convergent and
can be calculated using standard methods.
In the computation of ${\cal K}_j(0)$ we integrate first over
frequency without any cutoff. This integral is still
finite. After this step, we introduce an isotropic gaussian
cutoff for the space directions of order $\Lambda \approx \frac{1}{a_0}$
where $a_0$ is the lattice spacing. However, once the integration
over frequency has been performed, the finite contribution of the
divergent term ${\cal K}_j(0)$ is independent of the particular form
of the cutoff being used.

The computation of ${\cal K}_j(0)$ with $j=0,1,2,3$ is nothing
but the calculation of the critical coupling constants
performed in section\ (\ref{sec:mft}).
We have
\begin{mathletters}
\label{tuitas}
\begin{equation}
{\cal K}_3(0) =
 i{\rm tr}\left[{\bf {\hat S}}_L(\vec k)\tau_3{\bf {\hat S}}_U(\vec k)
\tau_3\right]
=
\frac{1}{2\sqrt{\pi}}  \Lambda - \frac{1}{2\pi} m\left( 1 + s \right)
\end{equation}
\begin{equation}
{\cal K}_0(0)  =
i{\rm tr}\left[{\bf {\hat S}}_L(\vec k)
\gamma_0{\bf {\hat S}}_U(\vec k)\gamma_0\right]
=
\frac{1}{2\pi} m \left( 1 - s \right)
\end{equation}
\begin{equation}
{\cal K}_j(0) =
i{\rm tr}\left[{\bf {\hat S}}_L(\vec k)
\gamma_j\tau_j{\bf {\hat S}}_U(\vec k)\gamma_j\tau_j\right]
=
-\frac{1}{4\sqrt{\pi}} \Lambda - \frac{1}{2\pi} m \left( 1 - s \right)
\end{equation}
\end{mathletters}
where $j=1,2$.
To get ${\cal K}_3(\vec q) - {\cal K}_3(0)$ we need to integrate
\begin{equation}
{\cal I}_3^E = - 4 \int {dk^3 \over (2\pi)^3}
\frac{(k^2 + m^2) \vec q\cdot(\vec q-\vec k) 
+ m^2(1 + s)\vec q\cdot(2 \vec k -\vec q)}{
(k^2 + m^2)^2 ((k-q)^2+m^2)}
\label{scaker}
\end{equation}
where $k^2$, $q^2$, $\vec k$ and $\vec q$ refer to the
imaginary frequency rotated form of
the tri-vectors $k_\mu$ and $q_\mu$, and $s = {\rm sgn}(m_L)
\cdot{\rm sgn}(m_U)$.

For the frequency channel given by $\varphi_0$ we get
after rotating to imaginary frequency
\begin{equation}
{\cal I}_0^E = - 4 \int_{k}
\left\{
\frac{\left( 2 k_0 q_0 + \vec q\cdot(\vec k-\vec q)\right) (k^2+m^2) -
\vec q\cdot(2\vec k-\vec q)\left( 2 k_3^2 + m^2 (1-s)\right)}
{(k^2+m^2)^2((k-q)^2+m^2)}
\right\}
\label{vecker}
\end{equation}
For the channels given by $\varphi_1$ and $\varphi_2$ we
obtain expressions similar to the one for $\varphi_0$ with
$k_0$ exchanged by $k_1$ or $k_2$ in each case.
The kernel for the spatial channels also have an
opposite sign to ${\cal K}_0(\vec q) - {\cal K}_0(0)$.

We may write down the denominators in Eq.\ (\ref{scaker}) and
Eq.\ (\ref{vecker}) in the form
\begin{equation}
{1\over D_E\left[ m^2 \right]}
=  \int_{0}^{1} du (1 - u) \int_{0}^{\infty} {\lambda}^{\ 2} d\lambda
           e^{-\lambda \vec l^{\ 2}}
          e^{-\lambda\left[ m^2 + u(1-u)\vec q^{\ 2}\right]}
\label{denom}
\end{equation}
In Eq.\ (\ref{denom}) we performed the change of variables
$\lambda=\alpha+\beta$ and $u=\beta/(\alpha+\beta)$,
we defined $\vec l \equiv \vec k - u\vec q$,
and once again, for
simplicity, we restricted to the case $|m_L|=|m_U|=m$.
After integrating over $k$ (or rather, $l$)
and $\lambda$ and a simple change of variables,
Eq.\ (\ref{scaker}) becomes
\begin{equation}
{\cal I}_3^E  = 
\frac{1}{2\pi}
\left\{
m(1+s) - \frac{2}{|\vec q|}\sin^{-1}\left( \frac{|\vec q}
{\sqrt{4 m^2 + \vec q^{\ 2}}}\right)
\left( m^2 (1+s) + \frac{1}{2} \vec q^{\ 2}\right)
\right\}
\label{ie3}
\end{equation}
For the vector channels we get
\begin{eqnarray}
{\cal I}_j^E  
& = &
\frac{1}{2\pi}g_{jj} 
\left\{ 
m
\left( 
s +2\kappa_j^2+3\kappa_j^4
\right)
+
\frac{2 m \vec q^{\ 2}}{4 m^2 +\vec q^{\ 2}}
\left( \kappa_j^4-\kappa_j^2\right)
\right\}
\nonumber \\
& - &
\frac{1}{2\pi}g_{jj}\left[ 2m^2 s + 4 m^2 
\left( \kappa_j^2 + \frac{3}{2}\kappa_j^4
\right) -\frac{\vec q^{\ 2}}{2}\left( 1 - 3\kappa^4\right)\right]
\frac{1}{|\vec q|}\sin^{-1}
\left( \frac{|\vec q|}{\sqrt{4 m^2 + \vec q^{\ 2}}}\right)
\label{iej}
\end{eqnarray}
In Eq.\ (\ref{iej}) $g_{jj}$ only indicates that the channels given
by $\varphi_1$ and $\varphi_2$ give a contribution with sign opposite
to the one given by $\varphi_0$. Also, we have defined
$\kappa_j \equiv\frac{q_j}{|\vec q|}$
and $ c\equiv 1 + \frac{4 m^2}{\vec q^{\ 2}}$.
The expressions given by Eq.\ (\ref{scalar}) to Eq.(\ref{vector})
can now be obtained simply by combining Eq.\ (\ref{tuitas})
with Eq.\ (\ref{ie3}) and Eq.\ (\ref{iej}).

\section{Gradient expansion}
\label{sec:C}

In order to obtain an effective theory valid for long wavelength 
excitations, in momentum space we only need an expansion to the
few lowest orders in the external momenta of the diagrams, since
each external momentum will generate a space derivative in the 
Fourier antitransformed expression.
Here we pursue further the expansion indicated by Eq.\ (\ref{lnexp}). 
However, instead of computing the loop diagrams exactly as we did in
order to calculate the correction to the ground state energy in 
section \ref{sec:order}, we expand up to second order in the external
momenta. Notice that we are not integrating out neither the gauge
fields nor the $\varphi$-fields in this case. 

\subsection{Diagrams with two external legs}
\label{subsec:twolegs}

We showed in section \ref{sec:order} that the classical energy is of
${\cal O}(N)$. From the expansion of the logarithm of the determinant,
to second order in power of the (small) fluctuating fields contained
in the operator $\hat{\bf Q}$ of section \ref{sec:order}
which now also includes the gauge fields in the planes --or to 
${\cal O}(1)$ in the $\frac{1}{N}$ expansion--, we have 
\begin{eqnarray}
i {1\over 2} \int dx^3
\left( \hat{\bf S}\hat{\bf Q}\hat{\bf S}\hat{\bf Q}\right)
& = & 
{i\over 2}
\int_q \int_k 
{\rm tr}\left[\hat{\bf S}_{L}(\vec k)
{\raise.15ex\hbox{$/$}\kern-.72em\hbox{$A$}}_L(\vec q)
\hat{\bf S}_{L}({\vec k} + {\vec q})
{\raise.15ex\hbox{$/$}\kern-.72em\hbox{$A$}}_L (-{\vec q})
\right]
\nonumber \\
& + &
{i\over 2}\int_q \int_k
{\rm tr}\left[\hat{\bf S}_{U}(\vec k)
{\raise.15ex\hbox{$/$}\kern-.72em\hbox{$A$}}_U(\vec q)
\hat{\bf S}_{U}({\vec k} + {\vec q})
{\raise.15ex\hbox{$/$}\kern-.72em\hbox{$A$}}_U(-{\vec q})
\right]
\nonumber \\
& + & 
2 \times
{i\over 2}
\int_q\int_k
{\rm tr}\left[\hat{\bf S}_{L}(\vec k)\hat{\bf \varphi}(\vec q)
\hat{\bf S}_{U}({\vec k} + {\vec q})
\hat{\bf \varphi}(-{\vec q})
\right]
\label{cuad}
\end{eqnarray}
As a shorthand we have used the notation
$ \int \frac{dk^3}{(2\pi)^3} \equiv \int_k$.

For the gauge fields alone, 
the diagram has an ultraviolet divergence that needs to be treated.
We use dimensional regularization 
to ensure transversality, {\it i.e.} to preserve gauge invariance.
The calculation is similar to the one shown in chapter VII of reference
\cite{book}.
The first two lines of the r.h.s. of Eq.\ (\ref{cuad}) \ give
\ 
$\int_q \Pi^{LL}_{\mu\nu}(q) A_{L}^{\mu}(q)
A_{L}^{\nu} (-{q})$
\
where
\begin{equation}
\Pi^{LL}_{\mu\nu} (\vec q) = i \ \epsilon_{\mu\nu\lambda} \ q^{\lambda} \ 
\Pi^{LL}_A (q^2) + 
\left( q_\mu q_\nu - q^2 g_{\mu\nu}\right)
\Pi^{LL}_S(q^2)
\label{pimunu}
\end{equation}
with
\begin{equation}
\Pi^{LL}_A (q^2) = \frac{1}{2\pi}\frac{m}{\sqrt{q^2}} \sinh^{-1}
\left( \frac{1}{\sqrt{\frac{4m^2}{q^2} - 1}}\right)
\end{equation}
and
\begin{equation}
\Pi^{LL}_S (q^2) = \frac{1}{8\pi} \ \frac{1}{\sqrt{q^2}}
\ \left[ - \frac{2|m|}{\sqrt{q^2}} +
\left( 1 + \frac{4 m^2}{q^2} \right) \sinh^{-1}
\left( \frac{1}{\sqrt{\frac{4m^2}{q^2} - 1}} \right)
\right]
\end{equation}
and the corresponding expression (UU) for the upper plane. 
This is the full expression for the polarization tensor for
the gauge fields.
The small momentum limit for Eq.\ (\ref{pimunu}) is given by
\[ \frac{1}{4\pi} \ \frac{m}{|m|} i \epsilon_{\mu\nu\lambda}
q^{\lambda} \ + \ \frac{1}{16\pi} \ \frac{1}{|m|}
\left( q_\mu q_\nu - q^2 g_{\mu\nu}\right) \]
Notice that the fermion mass with its sign enters
the antisymmetric part of $\Pi^{LL}_{\mu\nu}$.
Therefore, for an antisymmetric ordering of chiralities in the
ground state, $\Pi_A^{LL}$ and $\Pi_A^{UU}$ will bear opposite 
signs. 
In fact, the ratio $\frac{m}{|m|}$ in our case
is actually always positive $1$, since the sign of the masses have 
already been taken into account when definig $\hat{\bf S}_L$ and
$\hat{\bf S}_U$ in section \ref{sec:order}, and we are considering
the case when the magnitude of the masses is the same on both planes.
From now on, we give the details for the computation of the 1-loop diagrams
involving exclusively the scalar channel.
The third line in Eq.\ (\ref{cuad}) can be rewritten as
\begin{equation}
i \ \int_q \int_k   {\rm tr}  \left[ \hat{\bf S}_L (k) \ 
\tau_3 \ \hat{\bf S}_U (k + q) \ \tau_3 \right]
|\varphi (q)|^2
\end{equation}
This expression is similar to the ones we encountered in the previous 
appendix. 
Since we are interested in a small external momentum expansion,
the exact expression shown above can be approximated by
\begin{equation}
2 \ i \ \int_q \int_k  {\rm tr} 
\left[ \hat{\bf S}_L (k) \ \sum_{n=0}^{\infty} (-1)^n
\left( \hat{\bf S}_U (k) 
{\raise.15ex\hbox{$/$}\kern-.57em\hbox{$q$}}
\right)^n
\ \hat{\bf S}_U (k) \right]
|\varphi_3 (q)|^2
\end{equation}
which gives the following result
\begin{equation}
\int_q  |\varphi_3(q)|^2 \ 
\left[ \frac{1}{\pi} \ \left( \sqrt{\Lambda^2 + m^2} - \sqrt{m^2} \right)
+ \frac{1}{4\pi} \ \frac{1}{(m^2)^{1/2}} q_\mu \ q^{\mu} \right]
\end{equation}
The above expressions are valid
up to second order in the external momenta of the 1-loop diagram,
{\it i.e.}, in the momenta of the field $\varphi$ or in the 
momenta of the gauge fields.

\subsection{Diagrams with three external legs}
\label{subsec:threelegs}

The next term in the expansion of the logarithm of the fermionic
determinant is
\begin{eqnarray}
- \frac{1}{\sqrt{N}} & \ & \frac{i}{3} 
\int dx^3
\left( \hat{\bf S}\hat{\bf Q}\right)^{3}
 = \nonumber \\
 - \frac{3}{\sqrt{N}} & \times & \frac{i}{3} \int_p \int_q \int_k
{\rm Tr}\left[\hat{\bf S}_L (k) \gamma_\mu 
\hat{\bf S}_L({k} + {p})\hat{\bf \varphi}(q)
\hat{\bf S}_U ({k} + {p} + {q})
\left( \hat{\bf \varphi} ({p}+{q})\right)^{*}\right]
A_L^{\mu} (p)
\nonumber \\
- \frac{3}{\sqrt{N}} & \times & \frac{i}{3} \int_p \int_q \int_k
{\rm Tr}\left[\hat{\bf S}_U (k) \gamma_\mu 
\hat{\bf S}_U({k} + {p})\left( \hat{\bf \varphi}(- q)\right)^{*}
\hat{\bf S}_L ({k} + {p} + {q})
\hat{\bf \varphi} (- {p}- {q})\right]
A_U^{\mu} (p)
\nonumber \\
\label{triangle}
\end{eqnarray}
Both terms in the r.h.s. of Eq.\ (\ref{triangle}) are similar. 
I can easily be shown that the zeroth order term in external momenta
vanishes. For the scalar channel only, 
up to second order in the external momenta, the three-leg 1-loop 
diagrams give the following contribution
\begin{eqnarray}
& - &\ \frac{1}{\sqrt{N}} \ \frac{1}{4\pi} \ \frac{1}{|m|}
\int_{{q}, {s}, {p}}
\left( q_\mu - s_\mu \right) 
\left( A_L^{\mu} (p) - A_U^{\mu}(p)\right)
\delta \left( {p} + { s} + {p}\right)
\varphi(q) \left(\varphi(-s)\right)^{*} 
\nonumber \\
& & \quad + \frac{i}{\sqrt{N}} \ \frac{1}{8\pi}
\frac{1}{|m|^2}
\epsilon_{\mu\nu\lambda}
\int_{{q}, {s}, {p}}
\left( A_L^{\mu} (p) + A_U^{\mu}(p)\right)
\ p^{\nu} \ s^{\lambda}
\delta \left( {p} + {s} + {p}\right)
\varphi(q) \left(\varphi(- s)\right)^{*} 
\end{eqnarray}

\subsection{Diagrams with four external legs}
\label{subsec:fourlegs}

The fourth order in the expansion of the logarithm of the
fermionic determinant gives 
\begin{eqnarray}
&& \frac{i}{4N} 
\int dx^3
\left( \hat{\bf S}\hat{\bf Q}\right)^{4}
 = \nonumber \\
&& + \frac{i}{N} \int_{l, p, q, k}
{\rm Tr} \left[ 
\hat{\bf S}_L (k) \gamma_\mu 
\hat{\bf S}_L({k} + {p}) \gamma_\nu
\hat{\bf S}_L({k} + {p} + {l})
\hat{\bf \varphi}(q)
\hat{\bf S}_U ({k} - {s})
\left( \hat{\bf \varphi} (-{s})\right)^{*}\right]
A_L^{\mu} (p) \
A_L^{\mu} (l)
\nonumber \\
&& + \frac{i}{N} \int_{l, p, q, k}
{\rm Tr} \left[ 
\hat{\bf S}_U (k) \gamma_\mu 
\hat{\bf S}_U ({k} + {p}) \gamma_\nu
\hat{\bf S}_U ({k} + {p} + {l})
\hat{\bf \varphi}(q)
\hat{\bf S}_L ({k} - {s})
\left( \hat{\bf \varphi} (-{s})\right)^{*}\right]
A_U^{\mu} (p) \
A_U^{\mu} (l)
\nonumber \\
&& + \frac{i}{N} \int_{l, p, q, k}
{\rm Tr}\left[
\hat{\bf S}_L (k) \gamma_\mu 
\hat{\bf S}_L ({k} + {p})
\hat{\bf \varphi}(q)
\hat{\bf S}_U ({k} + {p} + {q})
\gamma_\nu
\hat{\bf S}_U ({k} - {s})
\left( \hat{\bf \varphi} (-{s})\right)^{*}\right]
A_L^{\mu} (p) \
A_U^{\mu} (l)
\nonumber \\
&& + \frac{i}{2N} \int_{l, p, q, k}
{\rm Tr}\left[
\hat{\bf S}_L (k) 
\hat{\varphi} (p)
\hat{\bf S}_U ({k} + {p})
\left(\hat{\bf \varphi} (q)\right)^{*}
\hat{\bf S}_L ({k} + {p} + {q})
\hat{\varphi}(l))
\hat{\bf S}_U ({k} - {s})
\left( \hat{\bf \varphi} (-{s})\right)^{*}\right]
\nonumber \\
&& + 
\ \ {\rm terms \ involving \ four \ gauge \ fields}
\label{cuadri}
\end{eqnarray}
From this expression we are going to consider only the four first lines
on the r.h.s. of Eq.\ (\ref{cuadri})
as they will show to be the relevant terms for our gradient expansion.
As we did before, we consider only 
the scalar channel.
The third term on the r.h.s. of Eq.\ (\ref{cuadri}) 
gives 
a total contribution, valid
to first order in the
external momenta that looks
\begin{eqnarray}
- \frac{1}{N} \ \frac{1}{16\pi} \ \frac{1}{(m^2)^{1/2}} \ 
\Big\{ 8 \  
g_{\mu\nu} \ 
+ \  \frac{m }{|m|} \ 
{\rm tr} \left[ \gamma_\mu \ \gamma_\nu \ \left(
{\raise.15ex\hbox{$/$}\kern-.57em\hbox{$p$}} 
+ 
{\raise.15ex\hbox{$/$}\kern-.57em\hbox{$l$}}
\right) \ \right]
\Big\} A_L^{\mu} (p) \ A_U^{\nu} (l)
\ \varphi_3(q) \left(\varphi_3(- s ) \right)^{*}.
\label{cuadriLU}
\end{eqnarray}
The contribution coming from the first term on the r.h.s.
of Eq. \ (\ref{cuadri}) is
\begin{eqnarray}
- \frac{1}{N} \ \frac{1}{16\pi} \ \frac{1}{(m^2)^{1/2}} 
\Big\{- 4 \ g_{\mu\nu} 
+ \frac{m }{|m|} \ 
{\rm tr} \left[ \gamma_\mu \ \gamma_\nu \ \left(
{\raise.15ex\hbox{$/$}\kern-.57em\hbox{$p$}}
+ 
{\raise.15ex\hbox{$/$}\kern-.57em\hbox{$q$}}
 \right) \ \right]
\Big\} A_L^{\mu} (p) \ A_L^{\nu} (l)
\ \varphi_3(q) \left(\varphi_3(- s ) \right)^{*}.
\label{cuadriLL}
\end{eqnarray}
The second term on the r.h.s. of Eq. \ (\ref{cuadri}) gives
\begin{eqnarray}
- \frac{1}{N} \ \frac{1}{16\pi} \ \frac{1}{(m^2)^{1/2}} 
\Big\{- 4 \ g_{\mu\nu} 
- \frac{m }{|m|} \ 
{\rm tr} \left[ \gamma_\mu \ \gamma_\nu \ \left(
{\raise.15ex\hbox{$/$}\kern-.57em\hbox{$p$}}
+ 
{\raise.15ex\hbox{$/$}\kern-.57em\hbox{$s$}}
\right) \ \right]
\Big\} A_U^{\mu} (p) \ A_U^{\nu} (l)
\ \varphi_3(q) \left(\varphi_3(- s ) \right)^{*}.
\label{cuadriUU}
\end{eqnarray}
The origin of the relative sign between the antisymmetric parts of 
Eq. \ (\ref{cuadriLL}) and Eq. \ (\ref{cuadriUU}) is the relative sign
of the fermion masses on the planes.


\newpage

\begin{references}

\bibitem{kla}
R.~B.~Laughlin, {\sl Science}~{\bf 242}, 525 (1988);
V.~Kalmeyer and R.~B.~Laughlin, {\sl Phys.~Rev.~Lett.~}{\bf 59}, 
2095 (1987).
\bibitem{wwz}
X.~G.~Wen, F.~Wilczek and A.~Zee, {\sl Phys.~Rev.\/}{\bf B39}, 
11413 (1989).
\bibitem{prediction} X.~G.~Wen and A.~Zee {\sl Phys.~Rev.~Lett.~} {\bf 62}, 2873
(1989);  {\sl Phys.~Rev.~}{\bf B43}, 5595, 1991.
\bibitem{experiments}
K.~B.~Lyons {\it et.~ al.\/}  {\sl Phys.~Rev.~Lett.~} {\bf 64}, 2949 (1990); 
S.~Spielman {\it et.~ al.\/} {\sl Phys.~Rev.~Lett.~} {\bf 65}, 123 (1990).
\bibitem{rojo}
A.~G.~Rojo and A.~J.~Leggett, {\sl Phys.~Rev.~Lett.~}{\bf 67}, 
3614 (1991).
\bibitem{lee}
M.~Ubbens and P.~A.~Lee, {\sl Phys.~Rev.~}{\bf B50},
438 (1994).
\bibitem{millis}
A.~J.~Millis and H.~Monien, {\sl Phys.~Rev.~Lett.~}{\bf 70},
2810 (1993); {\sl Phys.~Rev.~}{\bf B50}, 16606 (1994).
\bibitem{scalapino}
A.~W.~Sandvik and D.~J.~Scalapino, {\sl Phys.~Rev.~Lett.~}{\bf 72}, 2777 (1994).
\bibitem{gaitonde}
D.~M.~Gaitonde, D.~P.~Jatkar and S.~Rao, {\sl Phys.~Rev.~}{\bf B46}, 
12026 (1992).
\bibitem{afm}
I.~Affleck and J.~B.~Marston, {\sl Phys.~Rev.~}{\bf B37}, 3774 (1988).
\bibitem{kotliar}
G.~Kotliar, {\sl Phys.~Rev.~}{\bf B37}, 3664 (1988).
\bibitem{book}
E.~Fradkin, {\it ``Field Theories of Condensed Matter
Systems"}, Addison-Wesley, Redwood City (1991).
\bibitem{jackiw} R.~Jackiw, S.~Deser and S.~Templeton, 
{\sl Phys.~Rev.~Lett.~}{\bf 48},
372 (1982).
\bibitem{gross}
D.~J.~Gross and A.~Neveu, {\sl Phys.~Rev.~}{\bf D10}, 3235 (1974).
\bibitem{wiegmann}
P.~B.~Wiegmann, {\sl Phys.~Rev.~Lett.~}{\bf 60}, 821 (1988).
\bibitem{polyakov} A.~M.~Polyakov, {\sl Nucl.~Phys.\/} {\bf B120}, 429 (1977).
\bibitem{fidel} Eduardo Fradkin and 
Fidel Schaposnik, {\sl Phys.~Rev.~Lett.~}{\bf 66},
276 (1991).
\bibitem{thesis} C.~R.~Cassanello, University of Illinois Thesis, 1996.
\end{references}
\end{document}